\documentclass[%
reprint,
longbibliography,
nofootinbib,
amsmath,amssymb,
prd,
floatfix,
]{revtex4-2}

\usepackage{xcolor}
\usepackage{pifont}
\usepackage{graphicx}
\usepackage{dcolumn}
\usepackage{multirow}
\usepackage{bm}
\usepackage{hyperref}
\hypersetup{
  colorlinks   = true, 
  urlcolor     = blue, 
  linkcolor    = blue, 
  citecolor   = blue 
}

\AtBeginDocument{%
    \newwrite\bibnotes
    \def\bibnotesext{Notes.bib}
    \immediate\openout\bibnotes=\jobname\bibnotesext
    \immediate\write\bibnotes{@CONTROL{REVTEX42Control}}
    \immediate\write\bibnotes{@CONTROL{%
    apsrev42Control,author="48",editor="1",pages="1",title="0",year="1"%
    }%
    }
     \if@filesw
     \immediate\write\@auxout{\string\citation{apsrev42Control}}%
    \fi
}%

\begin{document}

\preprint{APS/123-QED}

\title{Future prospects on testing extensions to \texorpdfstring{$\Lambda$}{Lambda}CDM through the weak lensing of gravitational waves}%

\author{Charlie T. Mpetha$^{1}$}
\email{c.mpetha@ed.ac.uk}
\author{Giuseppe Congedo$^{1}$}%
\author{Andy Taylor$^{1}$}
\affiliation{%
$^{1}$Institute for Astronomy, School of Physics and Astronomy, University of Edinburgh,
Royal Observatory, Blackford Hill, Edinburgh, EH9 3HJ, United Kingdom}%

\date{\today}

\begin{abstract}
With planned space-based and 3$^{\rm rd}$ generation ground-based gravitational wave detectors (LISA, Einstein Telescope, Cosmic Explorer), and proposed DeciHz detectors (DECIGO, Big Bang Observer), it is timely to explore statistical cosmological tests that can be employed with the forthcoming plethora of data, $10^4-10^6$ mergers a year. We forecast the combination of the standard siren measurement with the weak lensing of gravitational waves from binary mergers. For 10 years of 3$^{\rm rd}$ generation detector runtime, this joint analysis will constrain the dark energy equation of state with marginalised $1\sigma$ uncertainties of $\sigma(w_0)\!\sim\!0.005$ and $\sigma(w_a)\!\sim\!0.04$. This is comparable to or better than forecasts for future galaxy/intensity mapping surveys, and better constraints are possible when combining these and other future probes with gravitational waves. We find that combining mergers with and without an electromagnetic counterpart helps break parameter degeneracies. Using DeciHz detectors in the post-LISA era, we demonstrate for the first time how merging binaries could achieve a precision on the sum of neutrino masses of $\sigma(\Sigma m_{\nu})\!\sim\!0.05\,$eV using $3\times10^6$ sources up to $z=3.5$ with a distance uncertainty of $1\%$, and $\sim\,$percent or sub-percent precision also on curvature, dark energy, and other parameters, independently from other probes. Finally, we demonstrate how the cosmology dependence in the redshift distribution of mergers can be exploited to improve dark energy constraints if the cosmic merger rate is known, instead of relying on measured distributions as is standard in cosmology. In the coming decades gravitational waves will become a formidable probe of both geometry and large scale structure.
\end{abstract}
\maketitle

\section{Introduction}

Gravitational Waves (GWs) are becoming a powerful cosmological tool. In 2017, coincident GW \cite{GW170817det} and electromagnetic (EM) \cite{GRB17B,170817-MM} observations of a merging binary neutron star system heralded the beginning of multimessenger GW cosmology. This event, named GW170817, was used to measure the present expansion rate of the Universe $H_0$. The associated gamma ray burst GRB170817A allowed localisation of the binary to its host galaxy, providing a redshift. This, combined with the distance to the source found using properties of the GW waveform, gave a value of $H_0 = 70.0^{+ 12.0}_{-8.0}\,{\rm km}\,{\rm s}^{-1}\,{\rm Mpc}^{-1}$ \cite{GW170817}. Precise, independent measurements of $H_0$ are pertinent due to the apparent tension between local Universe measurements of $H_0$ (such as Type Ia Supernovae \cite{SNIA_P} and lensed Quasars \cite{H0licow}) and the value inferred from the Cosmic Microwave Background \cite{Planck}. Just 50 multimessenger events similar to GW170817 would give us sufficient precision on $H_0$ to favour the early or late universe measurement \cite{LIGO2pct}. The planned improvement to the current LIGO-Virgo-Kagra (LVK) network \cite{LIGO_FUTURE} makes this goal highly achievable within the next few years, motivating exploring the further possibilities of using GWs, beyond a standard siren measurement of $H_0$. By the 2040's, several other GW detectors (Einstein Telescope \cite{ET}, Cosmic Explorer \cite{CE}, LISA \cite{LISA}), both ground- and spaced-based, will be either taking data or in the later stages of their development. The combination of these detectors will allow an exploration of more GW frequency ranges (source types) than LVK, and with much improved sensitivity. This will dramatically increase the scope of GWs for astrophysics and cosmology. Large numbers of GWs from binary systems will be ideal for statistical cosmological tests. This study aims to evaluate the prospects of the weak lensing of gravitational waves, where correlations in the small perturbations of the GW propagation can be used to infer properties of the intervening matter distribution. Such observations open up the possibility of using GWs as a novel probe of not only the geometry of the Universe, but also of large scale structure.

There are a few advantages of performing a weak lensing analysis using GWs instead of galaxies, besides it being a novel probe. It is a very clean measurement, avoiding systematics such as intrinsic alignments or blending which plague present and future weak lensing studies \cite{WL_sys}. GW detectors will also observe mergers to much higher redshifts, up to $z\!\sim\!100$. As in the standard siren case, the limiting factor in this analysis is the redshift determination. If there was a reliable GW only redshift inference method, for example sufficient source numbers for a `Spectral Siren' analysis \cite{Spectral}, then properties of the large scale matter distribution could be probed up to very high redshifts. Although the lensing window function peaks at low redshift, direct observation of the `Dark Ages' would give us valuable observational information on structure formation, such that we do not need to extrapolate the properties of structure in this redshift regime from models. The key challenge of GW weak lensing compared to galaxy weak lensing is the difficulty of observing a GW and distinguishing it from detector noise.

The possibility for using the weak lensing of GWs (GW-WL) as a probe of cosmology was first discussed in Ref.\;\cite{ULTRA} and developed in Ref.\;\cite{GW_Cl_z_int}. GW-WL forecasting was performed in \citet{GCAT} (henceforth  \hyperlink{cite.GCAT}{CT19}), where it was demonstrated a joint analysis of GWs, utilising both the standard siren distance measurement and the matter field information encapsulated in their weak lensing, can break key degeneracies giving better constraints than would be obtained individually. WL is largely insensitive to the Hubble parameter $H_0$ \cite{HALL_H0}, while a  joint standard siren+GW-WL analysis is sensitive to the expansion rate. The standard siren measurement provides a tight constraint on geometry parameters such as the expansion rate $H_0$ and matter density fraction $\Omega_{\rm m}$, without relying on an external data set. In  \hyperlink{cite.GCAT}{CT19} the authors assumed a number density of GW sources of $1\,$deg$^{-2}$ ($15\,000$ sources), and that each was a well localised event similar to GW170817 (a binary neutron star merger with an EM counterpart). But the performance of future detectors is currently speculative, and highly dependent on the unknown distribution and number of binary mergers in the Universe. To fully explore the potential of cosmological constraints through a joint analysis of standard sirens and GW-WL, a range of assumptions of source properties is needed. The analysis should also be extended to varying models of dark energy and a varying neutrino mass --- parameters of interest over the coming decades.

Here we build upon the work of  \hyperlink{cite.GCAT}{CT19}. Tomographic weak lensing is used to exploit all available cosmological information. A variety of assumptions on source and detector properties are made, and we extend to more cosmological models to explore how GWs can be used to answer present problems in cosmology. We also include sources without an associated EM counterpart, which will comprise the bulk of future GW detections. We find that combining two merger populations (sources with and without an EM counterpart) helps break parameter degeneracies and improve constraints in the $w_0-w_a$ and $\Omega_m-S_8$ planes, where the clustering parameter $S_8 = \sigma_8 \sqrt{\Omega_{\rm m} / 0.3}$. We show how a cosmologically varying source redshift distribution in the weak lensing analysis can further improve dark energy constraints. This is an advantage of gravitational wave studies where an astrophysics motivated cosmic merger rate density is possible. For a combination of GW detectors in the 2040's with EM telescopes, we find a standard siren+weak lensing analysis will be capable of improving upon dark energy equation of state parameter constraints from galaxy surveys such as Euclid \cite{EUCLID} and the Vera C. Rubin Observatory \cite{LSST}, and intensity mapping surveys such as HIRAX \cite{HIRAX_inst}. As we extend further into the future, if DeciHz GW detectors such as DECIGO \cite{DECIGO} and the Big Bang Observer \cite{BBO} come to fruition then the weak lensing of GWs will not only provide ultra precise constraints on geometry parameters, but could even give us valuable information on structure formation \cite{high_angular} and constrain the sum of neutrino masses $\Sigma m_{\nu}$. This is the first demonstration of GWs from merging binaries being an independent probe on neutrinos. We show that, by combining GWs with redshifts from galaxy surveys, GW-WL will be competitive with other single probes such as The Vera C. Rubin Observatory \cite{LSST_mnu} and CMB-S4 \cite{CMBHD} in constraining $\Sigma m_{\nu}$. Even in the most cosmology agnostic model used, $\nu kw$CDM (curvature, dark energy and neutrinos all free parameters), we obtain percent or sub-percent level errors on all parameters, apart from $\Sigma m_{\nu}$. The clean and well understood measurement from GWs could be a valuable source of extra information when combining with other future surveys.

Section \ref{sec:SS} details future planned GW detectors, sources of GWs and their use for cosmology. In section \ref{sec:WLGW} the physics of the weak lensing of gravitational waves is introduced, before the modelling used in this analysis, including source assumptions and detector uncertainties, is described in section \ref{sec:model}. Finally the results are presented in section \ref{sec:res} and we conclude in section \ref{sec:concl}.


\section{Standard sirens: sources and future detectors}
\label{sec:SS}

Observing gravitational waves from compact object mergers provides the luminosity distance $d_L$ of the source, through determination of the amplitude and frequency evolution of the wave.  This measurement is model-independent, giving it a clear advantage over, for example, Type Ia Supernovae which depend on calibration of the cosmic distance ladder. This direct measurement of the distance from the waveform makes GWs `Standard Sirens'. The observed gravitational wave amplitude is given by the linear combination of the $+$ and $\times$ polarisation of the GW, combined with their corresponding antenna pattern functions $F$ describing a detectors response to the wave:
\begin{equation}
    h = F_{+}h_{+} + F_{\times}h_{\times} \, .
\end{equation}

To leading order, the individual components of the gravitational wave are given by
\begin{align}
    h_{+} &= \frac{4}{d_L}\left(\frac{G\mathcal{M}_{c}}{c^2}\right)^{5/3}\left(\frac{\pi f}{c}\right)^{2/3}\frac{(1+{\rm cos}^{2}\iota)}{2} \, {\rm cos}\left[\Phi(f)\right] \, , \\
    h_{\times} &= \frac{4}{d_L}\left(\frac{G\mathcal{M}_{c}}{c^2}\right)^{5/3}\left(\frac{\pi f}{c}\right)^{2/3}{\rm cos}\iota \, {\rm sin}\left[\Phi(f)\right] \, .
\end{align}
The speed of light is given by $c$. $\mathcal{M}_{c}$ is the redshifted chirp mass, $\mathcal{M}_{c} = (1+z)(M_{1}M_{2})^{3/5}/(M_{1}+M_{2})^{1/5}$ ($M_1$ and $M_1$ are the component masses), $f$ the frequency, $\iota$ is the inclination angle and $\Phi$ is the phase.
Observable binary mergers include binary neutron stars (BNS), a black hole and a neutron star (BHNS) and binary black holes (BBH), and the inspiral of these sources in quite universal. They can, however, be distinguished during the binary merger and ringdown through higher order terms in the Parameterised post-Newtonian formalism of the waveform. For more detail see Refs \cite{GW1_Magg,GW2_Magg}. 

The expansion rate $H(z)$ at redshift $z$ is related to $d_L$ by
\begin{equation}
        d_{L}(z) = (1+z)\int_{0}^{z} \frac{c \, dz'}{H(z')}\, , \label{eq:dL}
\end{equation}
where
\begin{align}
        \left(\frac{H(z)}{H_0}\right)^{2} &= \Omega_{\rm m}(1+z)^{3} + \Omega_{K}(1+z)^{2} \nonumber\\ &\quad + \Omega_{\rm DE}(1+z)^{3(1+w_0 +w_a)} e^{\frac{-3 w_a z}{1+z}} \, .
   \label{eq:Hz}
\end{align}
 $H_0$ is the present Universe expansion rate ($h = H_0 / 100\,{\rm km}\,{\rm s}^{-1}\,{\rm Mpc}^{-1}$ is the dimensionless expansion rate), $\Omega_{\rm m}$, $\Omega_{K}$ and $\Omega_{\rm DE}$ are the present matter, curvature and dark energy density parameters, while the widely used CPL parameterisation \cite{CPL1,CPL2} of a time-varying dark energy equation of state (EoS)
    \begin{equation}
        \frac{p_{\rm DE}}{\rho_{\rm DE}} \equiv w_{\rm DE}  = w_0 + w_a\frac{z}{1+z} \, ,
        \label{eq:w}
    \end{equation}
is assumed. $w_0$ is the dark energy EoS today, and $w_a$ is its growth rate with the scale factor $a$.
This functional form can accurately recreate the EoS for alternative dark energy models, such as a scalar field dark energy. Encapsulated in $\Omega_{\rm m}$ is the baryonic, cold dark matter and neutrino density components,
\begin{equation}
\Omega_{\rm m} = \Omega_{b} + \Omega_{\rm CDM} + \Omega_{\nu} \, .
\label{eq:om}
\end{equation}
In $\Lambda$CDM most of the matter in the Universe is Cold Dark Matter (CDM), dark energy is a cosmological constant $\Lambda$ ($\Omega_{\rm DE}=\Omega_{\Lambda}$) with a constant EoS of $w_{\Lambda}=-1$ ($w_0 =-1$ and $w_a = 0$) and there is zero curvature ($\Omega_{K}=0$). 

If the redshift of the GW source can also be determined, then a $d_L-z$ relation can be used to probe the cosmological parameters in Eqs\;(\ref{eq:dL}-\ref{eq:om}). This redshift determination is the limiting factor for cosmological applications of GWs. Due to the mass-redshift degeneracy in the GW waveform, some method of breaking this degeneracy, or external data, is needed to determine the redshift. GW sources can be broadly split into two categories, depending on whether they can or can not be localised to their host galaxy. These two categories are Bright Standard Sirens (BSS) and Dark Standard Sirens (DSS). BSS represent an ideal case for cosmology. Either an associated EM counterpart, or very precise sky localisation due to a high SNR, allows an accurate source redshift determination \cite{Schutz}. For multimessenger events, there is the possibility to probe deviations from GR through comparing the EM and GW propagation \cite{GWs_GR}. In the case of DSS, there is no associated electromagnetic counterpart and the sky localisation is too poor to identify the host galaxy. These sources make up the bulk of gravitational wave detections, therefore statistical redshift inference methods for large numbers of DSS are receiving more attention in the ramp up towards $3^{\rm rd}$ generation (3G) GW detectors.

BNS mergers may produce an observable EM counterpart, depending on the merger distance and inclination angle. Supermassive black hole (SMBH) and BHNS mergers may produce an EM counterpart, but for SMBH mergers their rate is too small for statistical cosmology applications \cite{LISA_SMBH}. For BHNS mergers their rate is uncertain, as is the fraction of these that will produce a detectable counterpart, though the expectation is they will mostly be DSS \cite{BHNS_DSS}. Most DSS are stellar binary black hole (BBH) mergers, and for these several redshift inference methods have been proposed in the absence of a counterpart. These include: a statistical average of galaxies within the GW localisation error box \cite{Schutz,GWXGAL,statz_2ndGen} which has already been performed on LVK data \cite{LIGO_DSS}, correlating the clustering of GW sources with the clustering of galaxies \cite{SSXSN,Mukherjee_2020} or another tracer such as HI intensity mapping \cite{GW_HI_XC}, by breaking the mass-redshift degeneracy using information on the source frame mass distribution \cite{LIGO_PISN,SFMD,Spectral}, tidal corrections in the late-inspiral signal of a BNS merger \cite{BNS_EOS_z}, or an expected cosmic merger rate density of sources \cite{Noon,DarkCosmo}. 

The so-called 3G ground-based GW detectors expected during the 2030's-2040's include the triangular configuration Einstein Telescope (ET), and Cosmic Explorer (CE) which is similar to LVK with longer arm lengths. In the same time period we will see the first space-based GW detector, LISA. The combined ground-based detectors will observe in the frequency range $1-10^{3}$ Hz and be sensitive to most BBH and BNS mergers in the Universe \cite{CE2,SFMD}, observing $\mathcal{O}(10^{4}-10^{6})$ binary mergers every year \cite{LISTEN,GWFAST}. The large uncertainty is due to the uncertain cosmic merger rates for different binary populations (see e.g \cite{Rates_Summary} for a summary). LISA, observing in the $10^{-4}-1$ Hz frequency range, will observe SMBH inspirals, stellar binaries before their inspiral and merger is detected by ground-based observatories \cite{multiband}, as well as many extreme mass-ratio inspirals (EMRI), which can also be used for cosmology \cite{LISA_EMRI}. Space-based GW detectors observing in the DeciHz regime, such as DECIGO \cite{DECIGO} and Big Bang Observer (BBO) \cite{BBO}, are possible successors to LISA and will bridge the gap between these two frequency ranges. Their design is far more ambitious, with four LISA-like constellations (with smaller arm lengths) in heliocentric orbits, two of which at the same location and the other two distributed around the Sun. They could observe the inspiral of binaries over periods of months before they enter the ground-based frequency range during their merger. This, coupled with their changing position and orientation, allows observations of similar numbers of mergers to 3G detectors but with much improved sky localisation, sufficient to localise a merger to its host galaxy even in the absence of a transient EM counterpart. For example BBO could contribute $\mathcal{O}(10^{5})$ highly localised ($\sim\!$ few arcsec) binary mergers per year \cite{ULTRA}. But while 3G ground-based detectors such as ET and CE are in the later stages of their development, the difficulty of space missions makes the future of detectors such as DECIGO and BBO less certain, and strongly dependent on the success of LISA.

To summarise, from 3G detectors we can expect $\mathcal{O}(10^{5}-10^{7})$ sources observed by ground-based observatories by the end of the 2040's, and comparable numbers with much improved sky localisation in the second half of the century should space-based DeciHz detectors be launched.

\section{\label{sec:WLGW} Weak lensing of gravitational waves}

The gravitational potential that is created by large-scale structure in the Universe induces small perturbations in the path of propagating radiation. The radiation is weakly lensed, and the resulting magnification of both electromagnetic \cite{SCHNEIDER,WL_B+S} and gravitational \cite{WL_GW_kappa} radiation in the geometric optics regime is given by
\begin{equation}
    \mu \approx 1 + 2\kappa \, ,
\end{equation}
where $\kappa$ is the lensing convergence, a weighted projection of density perturbations along the line-of-sight.

\begin{figure}
    \centering
    \includegraphics[width=8.6cm]{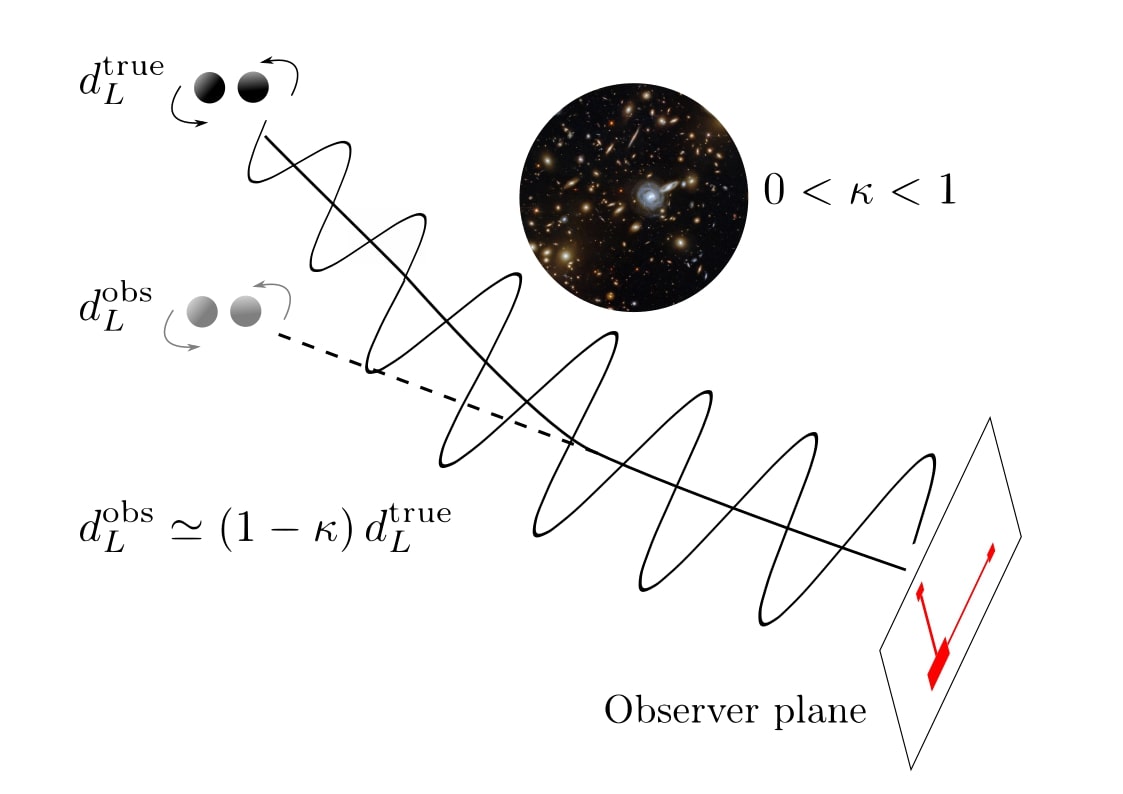}
    \caption{Gravitational waves being lensed by a large matter overdensity. The magnification results in a larger wave amplitude, and a smaller inferred distance to the source. The size of the deflection and change in amplitude has been exaggerated for illustration purposes.}
    \label{fig:wl}
\end{figure}

For GWs the situation is seen in Fig.\;\ref{fig:wl}. The amplitude of the wave is increased as the wave is lensed by an overdensity, leading to a smaller inferred luminosity distance. By relating the lensed and unlensed fluxes we obtain an expression for the observed, lensed luminosity distance $d^{\rm obs}_L$ in terms of the true $d_L$ and the convergence:
\begin{equation}
    d^{\rm obs}_{L} = \frac{1}{\sqrt{\mu}} d_{L}^{\rm true} \simeq (1 - \kappa) d_{L}^{\rm true}\, . \label{eq:dL_WL}
\end{equation}
There is also a small modification to the gravitational wave phase. In the geometric optics limit any phase fluctuations will just correspond to an arrival time shift making them observationally irrelevant in the case of weak lensing \cite{WL_GW_h_phase}. The strong lensing of gravitational waves, in the wave optics regime, is another science application of lensed GWs. In this case, it is useful for constraining properties of the lens and source populations \cite{SL-GW}, or as a further test of geometry \cite{SL-GW-H0}. To use GWs to explore the large scale matter distribution we turn to their weak lensing.

From a set of many mergers, a best fit $d_L -z$ relation can be constructed, sensitive to cosmological parameters. This is the standard siren measurement. The weak lensing of each source will create scatter around this best fit curve of order $1\%$, and due to the cosmological principle an average from large numbers of sources will recover the true relationship. Hence from this scatter, each source provides a point estimate of the convergence field. Though the lensing error dominates towards higher redshifts, uncertainties introduced by the source redshift, distance measurement and peculiar velocities of the sources make this is a noisy estimate of the convergence field.

Sources, in our case binary mergers, are binned in redshift intervals. The variance of $\kappa$ at different multipole modes $\ell$ (the Fourier conjugate of the angular separation) across tomographic redshift bins $i$ and $j$ is found through a two-point correlation function of the Fourier transformed convergence field in each bin,
\begin{equation}
    \langle\tilde{\kappa}_{(i)}(\bm{\ell})\tilde{\kappa}^{*}_{(j)}(\bm{\ell'})\rangle = \left(2\pi\right)^{2}\delta_{D}(\bm{\ell}-\bm{\ell'})\mathcal{C}^{\kappa\kappa}_{(ij)}(\ell) \, ,
    \label{eq:2PCF}
\end{equation}
where $\mathcal{C}^{\kappa\kappa}_{(ij)}(\ell)$ is the convergence power spectrum. This power spectrum is sensitive to cosmological parameters, and provides a complementary probe of information to the standard sirens, through a different filtering of the same data. The auto-correlation for each redshift bin  corresponds to $i=j$. Cross-correlating between bins adds information on the time variation of cosmological parameters. The convergence power spectrum between the $i^{\rm th}$ and $j^{\rm th}$ redshift bin is given by
\begin{align}
\mathcal{C}^{\kappa\kappa}_{(ij)}(\ell) &= \int_{0}^{z_{\rm max}} \frac{c\left(1+z\right)^{2}}{H(z)}\mathcal{W}_{(i)}(z)\mathcal{W}_{(j)}(z)  \nonumber \\& \qquad \qquad  \times  P_{\delta}\left(k=\frac{\ell+1/2}{\mathcal{K}(r)},z\right) \, dz \, ,
\label{eq:Cl} 
\end{align}
where
\begin{equation}
\mathcal{W}_{(i)}(z) = \frac{3}{2}\Omega_{\rm m,0}\left(\frac{H_{0}}{c}\right)^{2} \int_{z_{i}}^{z_{i+1}} p(z')\frac{\mathcal{K}(r'-r)}{\mathcal{K}(r')}\,dz' \, 
\end{equation}
is the Window function, containing the normalised observed source distribution $p(z)$ and the effect of the lens on this source due to their separation. The comoving distance $r$ is related to the transverse comoving distance $\mathcal{K}(r)$ by
    \begin{equation}
    \mathcal{K}(r) = 
\begin{cases}
    \frac{1}{\sqrt{\mid K\mid}}\sinh(\sqrt{\mid K\mid}r) \, ,& K < 0\, ,\\
    r\, ,              & K = 0 \, ,\\
    \frac{1}{\sqrt{\mid K\mid}}\sin(\sqrt{\mid K\mid}r)\, ,& K > 0\, ,
    \label{eq:TC}
\end{cases}
\end{equation}
where $K$ is the curvature.

The Limber approximation for $k$-modes, $k=(\ell+1/2)/\mathcal{K}(r)$, is shown in the matter power spectrum $P_{\delta}(k,z)$, and is used for large values of $\ell$. A common parameter to normalise the matter power spectrum is the linear normalisation of its amplitude in spheres of $8\,h^{-1}\,{\rm Mpc}$, $\sigma_8$ \cite{Peebles}, 
\begin{equation}
    \sigma^{2}_{8} = \sigma^{2}_{R=8\,h^{-1}\,{\rm Mpc}} = \int \frac{dk}{k}\frac{k^{3}P_{\delta}(k,z)}{2\pi^{2}}W^{2}(k, R) \, ,
\end{equation}
where $W(k,R)$ is a spherical tophat window function.

\section{Modelling Binary Merger Observations}
      
\label{sec:model}

\subsection{\label{sec:pz}Properties of binary merger populations}

In this study we use certain assumptions on the properties of the BSS and DSS, summarised by the following.
\begin{itemize}
    \item[---] For 3G detectors, all BSS events are assumed to be from merging binary neutron stars (BNS), while DSS are from both binary black hole (BBH) and black hole neutron star (BHNS) mergers. Though many observed BNS will also be DSS, BBH and BHNS will be the dominant contributors. For DeciHz detectors we assume well-localised BBH and BHNS mergers also contribute to the BSS sample. This is because for these detectors the constellations of satellites distributed around the Sun provide excellent source sky localisation \cite{BBO}, allowing many GWs to be localised to a host galaxy from which a redshift can be determined. 
    \item[---]  The redshift uncertainty for BSS,
    \begin{equation}
        \sigma_{z, {\rm BSS}} = \sigma_{0,{\rm BSS}} (1 + z)\, ,
    \end{equation}
    is spectroscopic, with $\sigma_{0,{\rm BSS}} = 0.001$ \cite{GW170817}. 
    \item[---] For DSS, a localisation uncertainty of $1\,$deg$^2$ is adopted \cite{BBH_error_Fisher,BBH_error_estimte}. The redshift is determined by a statistical average of galaxies in the localisation error box. We set         
    \begin{equation}
        \sigma_{z, {\rm DSS}} = \frac{\sigma_{0,{\rm DSS}}}{\sqrt{f(z) \, g(z)}} (1 + z) \, .
        \label{eq:DSS_zerr}
    \end{equation}
    Here, $\sigma_{0,{\rm DSS}} = 0.03$ representing expected photometric uncertainties from the Vera C. Rubin Observatory \cite{LSST_sigz}. Future high redshift spectroscopic galaxy surveys, such as MegaMapper \cite{MM}, could allow a statistical DSS method with spectroscopic redshifts (spec$-z$). The two functions $f(z)$ and $g(z)$ represent the GW detector and galaxy survey selection functions respectively, and are discussed further below.
    \item[---] BSS events are observed up to $z_{\rm max} = 2$ based on 3G GW detector horizons \cite{ET_BNS,CE} and future spectroscopic galaxy surveys \cite{WFIRST_specz,EUCLID}. 
    \item[---] For DSS, $z_{\rm max} = 3.5$, from the limit of future photometric galaxy surveys \cite{LSST_photometry}. These galaxies are needed to estimate the source redshift.
    \item[---] There are typically expected to be a factor of $10$ more BBH than BNS detections \cite{GW_CROSS,ET,LISA_BNS,LISA_STELLAR}. Only a fraction of BNS mergers will have EM counterparts. From this we assume $N_{\rm DSS}/N_{\rm BSS} = 100$ for 3G detectors, consistent with current LVK detections \cite{LIGO_Rates}.
\end{itemize}

The observed redshift distribution of binary mergers $p(z)$ is a combination of different factors,
\begin{equation}
    p(z) = p^{\rm{(th)}}(z) \, g(z) \, f(z) \, .
    \label{eq:pobs}
\end{equation}
For a given population of binaries, $p^{\rm{(th)}}(z)$ is a theoretical probability distribution derived from a theoretical cosmic merger rate density $\mathcal{R}(z)$, which can be modelled from the star formation rate and binary synthesis properties. The source number distribution is found from 
\begin{equation}
    n^{\rm{(th)}}(z) = 4\pi \frac{\mathcal{R}(z)}{(1+z)}\frac{dV_{c}}{dz} = 4\pi c \frac{\mathcal{R}(z)}{(1+z)}\frac{r^{2}(z)}{H(z)} \, ,
    \label{eq:pth}
\end{equation} 
where $V_{c}$ is the comoving volume. We use the \texttt{BPASS} predictions for transients \cite{BPASS}, which provides $\mathcal{R}(z)$ for different binary populations separately, including BNS, BBH and BHNS. This number distribution is then normalised to find the probability distribution,
\begin{equation}
    p^{\rm{(th)}}(z) = \frac{n^{\rm{(th)}}(z)}{\int_{0}^{z} n^{\rm{(th)}}(z')\,dz'} \, .
\end{equation}

Depending on GW source and detector parameters such as the sensitivity curve, antenna patterns, inclination angle, compact object masses etc., the GW may or may not be detected. This is modelled by the selection function $g(z)$. In \citet{pz_dark}, using a Fisher matrix analysis and SNR limit of 8 for merging BBHs, the authors find for a choice of 3G detector,
\begin{equation}
    g(z) \propto e^{-\left(r(z)/r_{\rm cut}\right)^{3}} \label{eq:gz} \, ,
\end{equation}
where the value $r_{\rm cut}$ for BBH mergers is given as $7.9\,$Gpc, which we adopt for the DSS case. This form is found by fitting a distribution of realistic sources that are above the SNR threshold as a function of comoving distance. In the BSS case a value of $r_{{\rm cut}}=4\,$Gpc is found by assuming the same functional form, but with a detector redshift limit of $z=2$, as will be the case for BNS observations.

The inclusion of the galaxy survey selection function $f(z)$ in Eq.\;(\ref{eq:pobs}) is important as we are assuming the redshift determination of merging compact objects is through galaxy observations, either from host galaxy determination, or from a statistical average of galaxies in a photometric galaxy survey. The inclusion of a source in the observed distribution depends on whether or not it has an associated redshift, hence depends on a relevant galaxy survey selection function. A simplistic choice has been made in this case, of a tophat function with a probability of 0.9 (completeness), then a smooth decay before reaching the maximum redshift. The shape is determined by a pivot redshift ($z_{\rm pivot} = 1.8$ for BSS \cite{EUCLID_design} and $2.8$ for DSS \cite{LSST_photometry}), 
\begin{align}
    f(z) &= \frac{0.9}{2}\left(1-{\rm tanh}\left(\frac{z-z_{\rm pivot}}{w\, z_{\rm pivot}}\right)\right) \, ,\label{eq:fz} \\
    w &= \frac{z_{\rm max}-z_{\rm pivot}}{2\,z_{\rm max}} \, .
\end{align}
These functions are also used to estimate the redshift uncertainty in the DSS case by acting as a shot noise scaling as shown in Eq.\;(\ref{eq:DSS_zerr}), recovering the expected behaviour of near-photometric uncertainties at low redshift \cite{LISA_EMRI,LISA_STELLAR}, and a rapid dilution to higher redshifts as the number of GW and galaxy observations decreases. The resulting $p^{(\rm obs)}(z)$ is shown for both BSS and DSS in Fig.\;\ref{fig:pz}. The sensitivity of the results to the forms of $g(z)$ and $f(z)$ is explored in Appendix \ref{app:sel}.
\begin{figure}
    \centering
    \includegraphics[width=8.6cm]{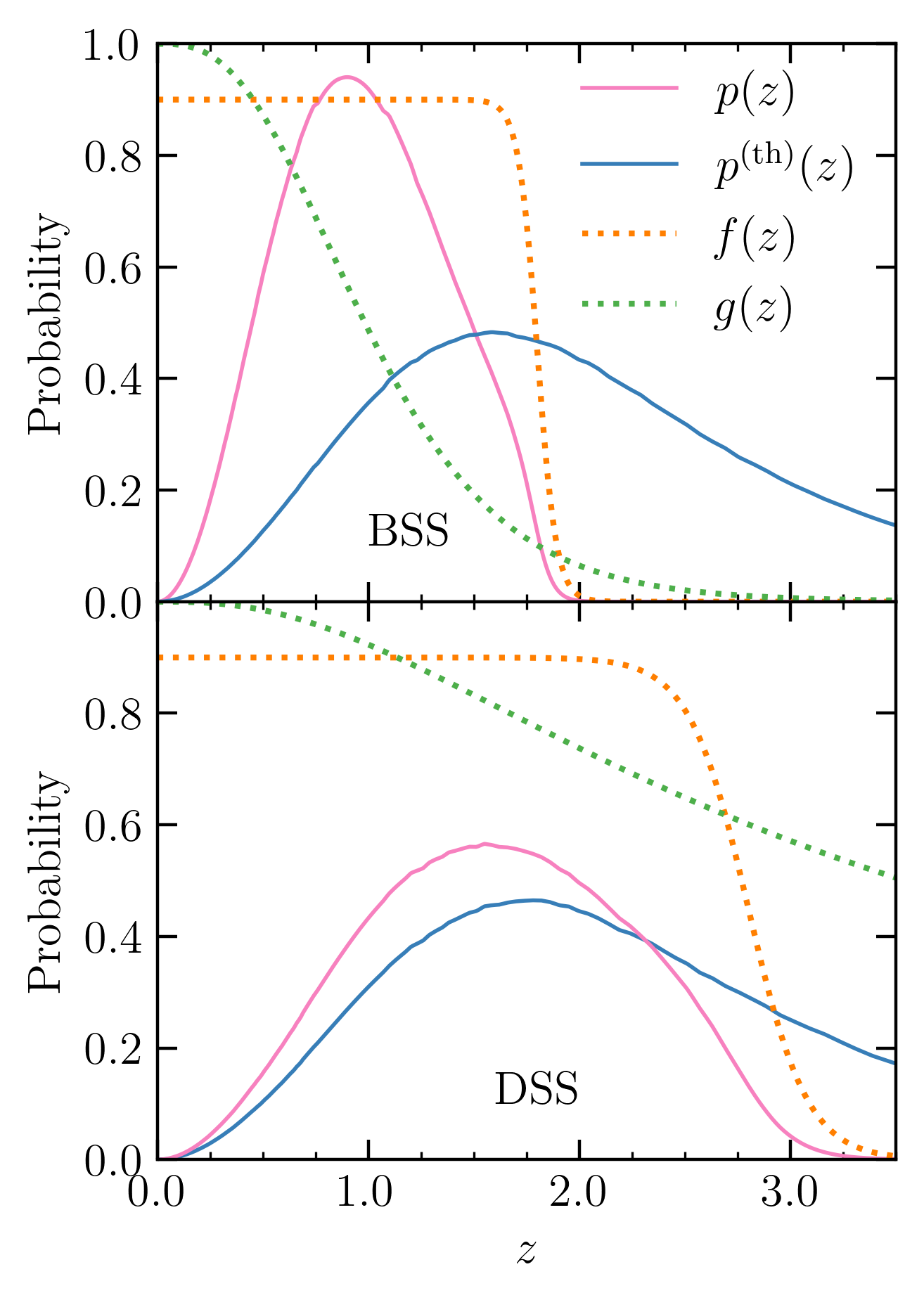}
    \caption{$p(z)$ is the observed redshift distribution of gravitational waves from binary mergers for 3G ground-based detectors. This is split into two populations; binary neutron stars with an EM counterpart (Bright Standard Sirens, top), and a combined population of binary black holes and black hole-neutron stars with a statistically obtained redshift (Dark Standard Sirens, bottom). $p(z)$ is obtained by combining the theoretical normalised redshift distribution $p^{\rm (th)}(z)$ (found from \texttt{BPASS} simulated merger rate densities \cite{BPASS}) with approximate forms of the relevant galaxy survey selection function $f(z)$ and $3^{\rm rd}$ generation gravitational wave detector selection function $g(z)$.}
    \label{fig:pz}
\end{figure}

\subsection{Source and detector uncertainties}
When using GWs as standard sirens, there are several sources of uncertainty. The GW detector instrumental uncertainty ($\sigma_{\rm GW}$) is assumed to have a range of values from $0.004 d_L$ to $0.1 d_L$. The source redshift uncertainty  ($\sigma_{z}$) is defined above. Uncertainty in the redshift owing to the source peculiar velocity  ($\sigma_{v_{ \rm pec}}$) is given the form
\begin{equation}
    \sigma_{v_{ \rm pec}} =  \frac{\sqrt{\smash[b]{\langle v_{\rm pec}^{2}\rangle}}}{c} (1 + z) \, .
\end{equation}
An rms value of $\sqrt{\smash[b]{\langle v_{\rm pec}^{2}\rangle}}\!\sim\!500\,{\rm km}\,{\rm s}^{-1}$ is adopted \cite{vpec500}. Finally is the uncertainty in the true value of $d_L$ due to the weak lensing of the GWs ($\sigma_{\rm WL}$). Different predictions for the size of $\sigma_{\rm WL}$ exist in the literature. A (pessimistic) fit used for Einstein Telescope forecasts \cite{0.05z_orig,ET_FISHER} assumes $5\%$ uncertainty at $z=1$ and linearly extrapolates: $\sigma_{\rm WL} / d_{L} = 0.05z$. A more appropriate expression derived in Ref. \cite{WL_error_derivation} using properties of the magnification probability distribution up to $z=3$ is given by
\begin{equation}
    \frac{\sigma_{\rm WL}}{d_{L}} = 0.066\left(\frac{1 - (1+z)^{-0.25}}{0.25}\right)^{1.8} \, .
    \label{eq:sig_WL}
\end{equation}

Other forms for $\sigma_{\rm WL}$ exist in the literature \cite{LENSINGERROR,Dancing}, Eq.\;(\ref{eq:sig_WL}) is used here as a middle estimate.
There has been development in `delensing' the GW, by observing the matter distribution along the line of sight of the GW source \cite{delens}. This can reduce the lensing uncertainty --- however it is not useful for our purposes, where we wish to extract information from the lensing of GWs. The combined uncertainty is
\begin{equation}
    \sigma_{d_L}^{2} = \sigma_{\rm GW}^{2} + \sigma_{\rm WL}^{2} + \left(\frac{\partial d_L}{\partial z}\right)^{2}\left(\sigma_{z}^{2} + \sigma_{v_{\rm pec}}^{2}\right) \, .
\end{equation}

The observed power spectrum  is the underlying convergence power spectrum, modified by a shot noise term due to our finite number of sources.
\begin{align}
    \mathcal{C}^{\rm obs}_{(ij)} &= \mathcal{C}^{\kappa\kappa}_{(ij)}(\ell) + \delta_{ij}\frac{\sigma_{\kappa}^{2}}{\bar{n}}e^{\left(\ell/\ell_{\rm max}\right)^{2}} \, , \label{eq:Clerror}\\
    \sigma_{\kappa}^{2} &= \sigma_{\rm GW}^{2} + \left(\frac{\partial d_L}{\partial z}\right)^{2}\left(\sigma_{z}^{2} + \sigma_{v_{\rm pec}}^{2}\right) \, ,
\end{align}
where $\bar{n}$ is the number density of GW sources, and $\delta_{ij}$ is the Kronecker delta ensuring shot noise only contributes for auto-correlations. Possible high-order correlations between the WL and GW signal are not included. To include the effects of sky localisation on the convergence power spectrum, we assume that a Gaussian kernel damps the signal. The total observed power spectrum is then deconvolved with the same kernel, leaving the Inverse-Gaussian term in the shot-noise error seen in Eq.\;(\ref{eq:Clerror}) \cite{Mukherjee_2020}. The maximum $\ell$ mode that can be probed is directly related to the localisation area, the uncertainty blows up past this $\ell$ mode. If the localisation area is given by $A_{\rm loc} = \Omega$ deg$^{2}$, then $\ell_{\rm max}\!\sim\!180/\sqrt{\pi \Omega}$.
    
Finally, the covariance is given by
\begin{align}
    &{\rm Cov}[\mathcal{C}^{\rm obs}_{(ij)}(\ell),\mathcal{C}^{\rm obs}_{(mn)}(\ell')] =  \nonumber \\ & \frac{\delta_{\ell\ell'}}{f_{\rm sky}(2\ell + 1)\Delta\ell}\left(\mathcal{C}^{\rm obs}_{(im)}(\ell)\mathcal{C}^{\rm obs}_{(jn)}(\ell)+\mathcal{C}^{\rm obs}_{(in)}(\ell)\mathcal{C}^{\rm obs}_{(jm)}(\ell)\right) \, ,
\end{align}
where trispectrum terms have been neglected. In this work we assume an observable sky area of $15\,000\,$deg$^{2}$ giving $f_{\rm sky}=0.36$ based on Euclid. This value considers the sky blocked from view by the galactic disk and zodiacal plane, and while GW observations are not limited by this, host galaxy redshifts are. 

\section{Forecasting a joint standard siren+weak lensing analysis \label{sec:res}}

\begin{table*}
\caption{\label{tab:parameters}
Parameters used in the Fisher forecasting analysis, their flat prior ranges, and the cosmological models in which they are treated as free (\checkmark) or set to constant (\ding{55}). $k\Lambda$CDM, $w$CDM and $\nu$CDM are $\Lambda$CDM with curvature, dark energy or neutrinos respectively as a free parameter. The most cosmology agnostic model in this analysis, $\nu kw$CDM, treats all these as free.}
\begin{ruledtabular}
\begin{tabular}{llllllll}
\multirow{2}{*}{\textrm{Parameter}}&
\multirow{2}{*}{\textrm{Fiducial}}&
\multirow{2}{*}{\textrm{Prior}}&
\multicolumn{5}{c}{Model}\\
 & & & $\Lambda$CDM & $k\Lambda$CDM & $w$CDM & $\nu$CDM & $\nu kw$CDM \\
\colrule
Dimensionless Hubble parameter $h$ & 0.673 & [0.5,0.9] & \checkmark  & \checkmark & \checkmark & \checkmark & \checkmark \\ 
Total matter density parameter $\Omega_{\rm m}$ & $0.316$ & $[0.1,0.9]$ & \checkmark &   \checkmark &\checkmark & \checkmark & \checkmark\\
Scalar spectral index $n_s$ & $0.965 $& $[0.5,1.5]$ & \checkmark & \checkmark &  \checkmark &\checkmark & \checkmark \\
Scalar amplitude ln$(10^{10} A_s)$ & $3.05$ & [2,4] & \checkmark & \checkmark & \checkmark & \checkmark & \checkmark  \\
Curvature density parameter  $\Omega_{\rm K}$ & $0$ & $[-0.3,0.3]$  & \ding{55}  & \checkmark & \ding{55} & \ding{55} & \checkmark \\
Dark energy EoS parameter $w_0$ & $-1$ & $[-2,0]$ & \ding{55}  & \ding{55} & \checkmark &\ding{55} & \checkmark\\
Dark energy EoS parameter $w_a$ & $0$ & $[-2,2]$ &\ding{55}  & \ding{55} &  \checkmark &\ding{55}& \checkmark\\
Sum of neutrino masses $\Sigma m_{\nu}$ [eV] & $0.06$ & $[0.005,1]$ & \ding{55}  & \ding{55}& \ding{55}& \checkmark & \checkmark\\
\end{tabular}
\end{ruledtabular}
\end{table*}

The real strength of this approach comes from the combination of the standard siren and weak lensing analyses, where the standard sirens essentially provide a strong constraint on geometry in the weak lensing analysis, without needing to rely on external datasets. The Bayesian framework is outlined in \hyperlink{cite.GCAT}{CT19}. Analytic derivatives can be found in the standard siren case, and numerical derivatives are needed when finding the Jacobian of the convergence power spectrum. Care is needed for these derivatives as the response of the power spectrum to a cosmological parameter can vary greatly over $\ell$-modes and redshift bins. Our solution is to, for each cosmological parameter, use the median value from the optimal steps of all modes and redshift bin correlations, then test it for optimality and stability. More detail can be found in Appendix \ref{app:derivatives}. A Fisher information matrix is constructed and used to find the Cram\'er-Rao lower bound on physical parameters, which the model depends on. These are then mapped to observed parameter uncertainties using Monte Carlo methods, as the mapping is often non-linear. We sum Fisher matrices to combine the results from the standard siren and weak lensing analyses, and all quoted constraints are $1\sigma$ marginalised uncertainties.

\texttt{Class} \cite{CLASS_cite} with \texttt{HMcode} \cite{HMcode} is used to compute $\sigma_8$ values and the auto- and cross- convergence power spectra between six redshift bins, where bin edges are defined such that there is an approximately equal number of sources per bin. A maximum $\ell$ mode of $\ell=3000$ is used to stay clear of the highly non-linear regime, where the statistical properties of the convergence field are close to Gaussian \cite{tomog}. Otherwise we could produce overly optimistic forecasts by assuming perfect knowledge of a regime with uncertain modelling. This is due to the importance of baryonic physics on such small scales \cite{baryon_model}. Nevertheless, there is still an assumption of much improved understanding of non-linearity in the matter power spectrum by the 2040's-2050's. We investigate the effect of decreasing the maximum $\ell$ mode on forecasts in Appendix \ref{app:lmax}. Fiducial parameter choices and the range of their flat priors are given in  Table~\ref{tab:parameters}, as well as a summary of the cosmological models used in this analysis, demonstrating which parameters are allowed to vary in which models. In this work we set $h^{2}\Omega_{b}=0.0224$ as a fixed parameter, motivated by its strong CMB constraint from the Planck satellite \cite{Planck}.

\subsection{2040's: 3G GW detectors}

In the 3rd generation of GW detectors, consisting of a network of ground-based interferometers such as the Einstein Telescope and Cosmic Explorer, there is expected to be up to $\sim\!10^{5}\,$yr$^{-1}$ observations of BNS mergers \cite{Lights_off} and $\sim\!10^{6}\,$yr$^{-1}$ for BBH mergers \cite{ET,CE,GWFAST}. In Ref.\;\cite{BNS_DETECTIONS} the authors forecast the expected number of BNS mergers with an associated gamma ray burst counterpart (making them BSS), using different assumptions on the GW and gamma ray detectors. An upper limit from CE+GECAM of $N\!\sim\!3\times10^{4}$ (translating to $\bar{n}\!\sim\!2\,$deg$^{-2}$ using our assumed fractional sky area of $f_{\rm sky} = 0.36$) over a 10 year runtime is found, and the number could be higher when considering other GW and EM detectors. The inclusion of the ET would give comparable or slightly higher numbers---there would likely be numerous coincident detections. Predictions for joint GW+EM observations (gamma ray bursts or X-rays) are also found in Refs\;\cite{MM_future,sirens_lure}, and are generally comparable or lower, agreeing with the range given in Ref.\;\cite{BNS_DETECTIONS}. These numbers are fairly small for statistical cosmological tests. In this case of relatively small numbers of BSS, then it is vital the abundant DSS detections are also used for cosmology. With a redshift determined from a statistical average of galaxies in the localisation error box of the GW, the redshift uncertainties of DSS are expected to be much larger than BSS. Despite this, the deeper redshift range and larger numbers means they can still be informative.

\begin{figure*}
    \centering
    \includegraphics[width=17.8cm]{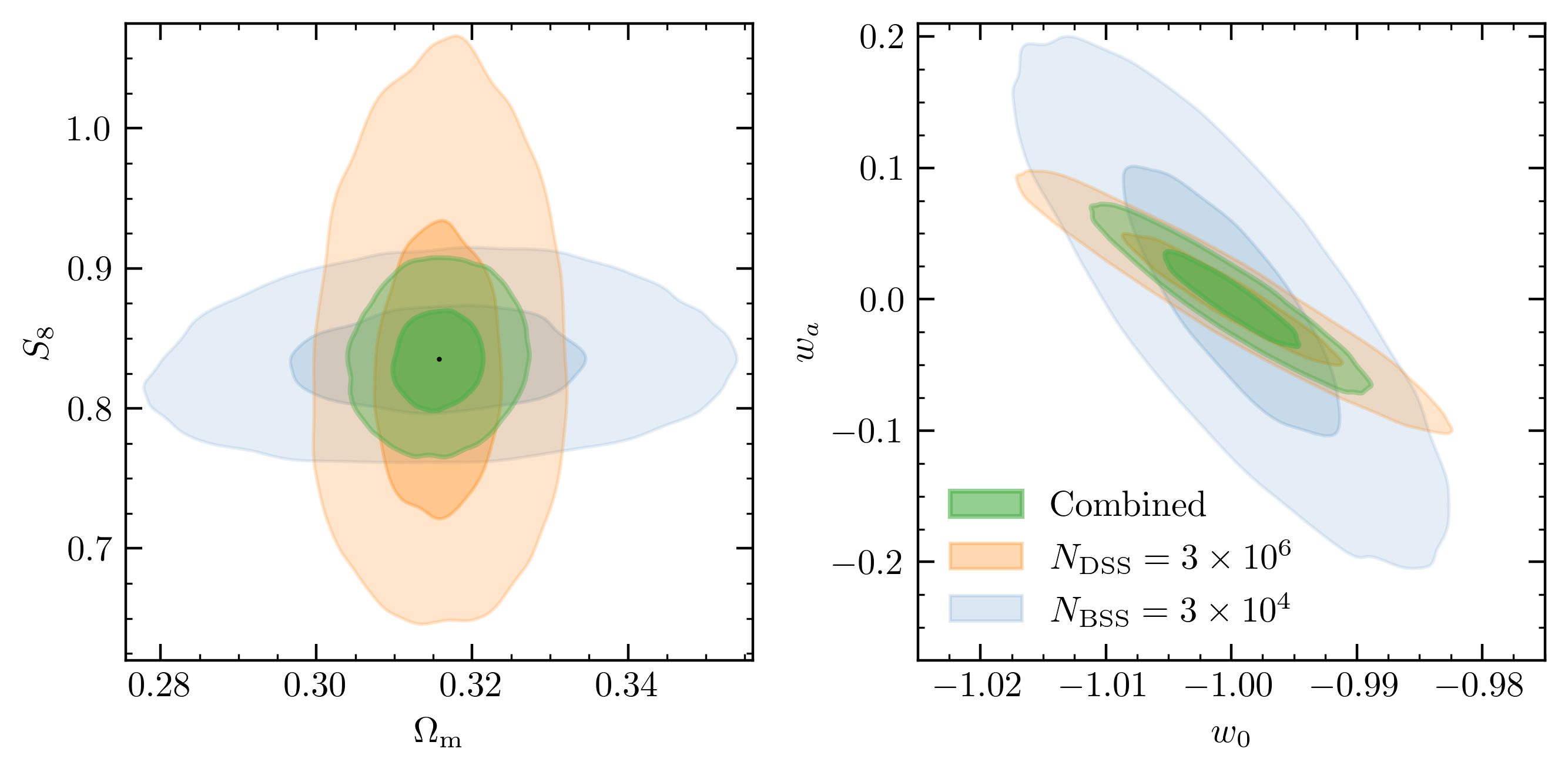}
    \caption{Forecasts for combined gravitational wave weak lensing+standard siren constraints on cosmological parameters in the $w$CDM model (see Tab.\;\ref{tab:parameters}) using realistic numbers for 3G detectors. Bright Standard Sirens (BSS) have a smaller redshift range but more accurate redshift determination through localisation to a host galaxy, while Dark Standard Sirens (DSS) can be observed up to larger redshifts, but their redshift is estimated using statistical methods and so is much noisier. We set the GW detector distance uncertainty to $2\%$ for BSS and $10\%$ for DSS.}
    \label{fig:wCDM_3G}
\end{figure*}

By the late 2030's, there may be other spectroscopic redshift surveys, for example the proposed MegaMapper \cite{MM} which would observe spec$-z$'s of $\sim\!10^8$ galaxies  from $2 < z < 5$. While not a benefit to spectroscopic follow up of BNS mergers, whose GW detector selection function drops off around $z=2$, MegaMapper could be used to perform statistical dark standard siren redshift estimation using spec$-z$'s. To test this possibility, we reran the analysis replacing the redshift uncertainty in Eq.\;(\ref{eq:DSS_zerr}) with $\sigma_{0,{\rm DSS}}=0.001$. This decreases the combined BSS+DSS parameter constraints by a factor of $0.5-0.8$.

\begin{figure*}
    \centering
    \includegraphics[width=17.8cm]{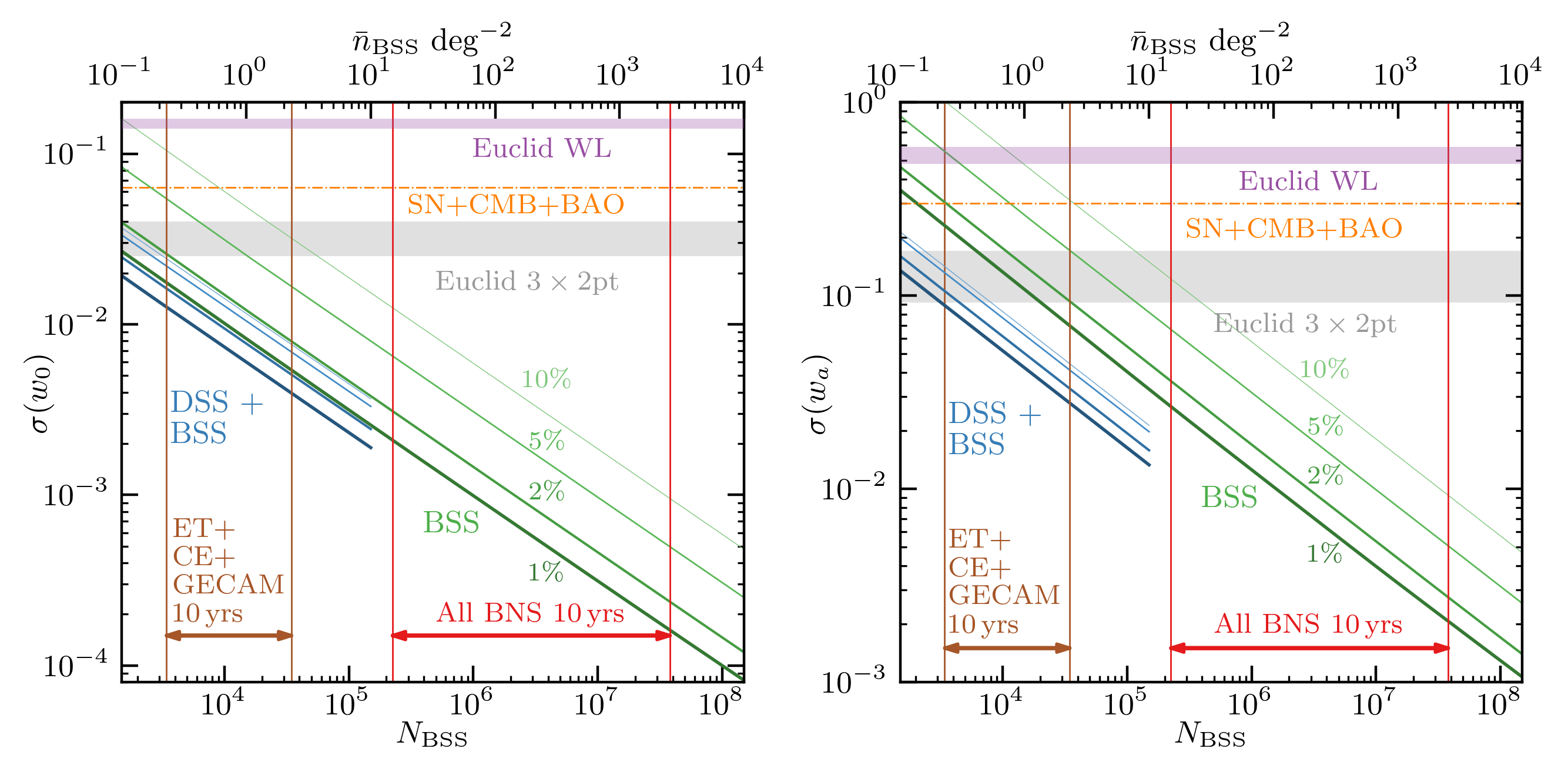}
    \caption{Forecasted errors in the $w$CDM model (see Tab.\;\ref{tab:parameters}) on the dark energy equation of state parameters $w_0$ (left) and $w_a$ (right) found from a joint gravitational wave weak lensing+standard siren analysis, as a function of the number (density) of Bright Standard Siren (BSS) observations. The green lines use a population of BSS with spectroscopic redshifts and a range of GW detector $d_L$ uncertainties, from $1\%-10\%$. The blue lines combine these populations with a set of Dark Standard Sirens (DSS) from binary black holes and black hole-neutron stars, with a statistical redshift determination and detector $d_L$ uncertainty of $10\%$. It is assumed that $N_{\rm DSS} = 100\,N_{\rm BSS}$. Horizontal bands give Euclid forecasts from either weak lensing (WL) alone, or the combined $3\times2$pt analysis. The orange dash-dot line is the most precise present constraint from Supernovae + CMB + BAO. The regions bound by vertical lines give expected numbers of BSS from the combination of 3G detectors Einstein Telescope (ET) and Cosmic Explorer (CE) with the gamma ray detector GECAM in brown, and a theoretical range for all BNS mergers in the Universe in red, both over 10 years of detector runtime.}
    \label{fig:w0wa}
\end{figure*}

Fig.\;\ref{fig:wCDM_3G} gives marginalised $1\sigma$ and $2\sigma$ contours for constraining the dark energy equation of state parameters $w_0$ and $w_a$, and matter density and clustering parameters $\Omega_{\rm m}$ and $S_8 = \sigma_8 \sqrt{\Omega_{\rm m} / 0.3}$ in the $w$CDM cosmological model (see Tab.\;\ref{tab:parameters}), using expected numbers after $10\,$years of 3G detector runtime. BSS are assumed to have an instrumental $d_L$ error $\sigma_{\rm GW} / d_L = 2\%$, while DSS have $\sigma_{\rm GW} / d_L = 10\%$ \cite{LISTEN}. This is motivated by the smaller number of well measured BSS, and the large number of DSS which will have a large range of instrumental uncertainties.  It can be seen how, due to the different degeneracy lines between BSS and DSS, their combination leads to improved constraints. This demonstrates the importance of including the large population of DSS to the sample. Despite their noisy redshift estimate, they probe higher redshifts leading to an improved $\Omega_{\rm m}$ and greater sensitivity to the time-variation of dark energy parameterised by $w_a$. But the larger sky localisation uncertainty means the $\mathcal{C}_{\ell}^{\kappa\kappa}$ result is degraded, hence the much weaker constraint on the clustering parameter $S_8$, which can only be probed through the weak lensing measurement of GWs. The dark energy equation of state parameters can be constrained through the standard siren measurement alone, and this provides the most information. Combining the standard siren measurement with weak lensing tomography, as is done for all results, improves constraints on geometry parameters by $\sim\!10\%$. Using 2D lensing (no tomographic bins) gives marginal added information for geometry parameters.

Fig.\;\ref{fig:w0wa} shows constraints for $w_0$ and $w_a$ in $w$CDM against source numbers, which is a proxy for detector runtime. Using the uncertainty on the LVK inferred local Universe BNS merger rate $\mathcal{R}(0)$ \cite{LIGO_Rates}, and assuming the same \texttt{BPASS} form for $\mathcal{R}(z)$, an estimate for the range of total mergers occurring over all cosmic history can be obtained, as an indication of future detection limits. This is done by normalising the \texttt{BPASS} $\mathcal{R}(z)$ to each upper and lower bound of $\mathcal{R}(0)$ in turn, substituting into Eq.\;(\ref{eq:pth}) and integrating over the redshift range. This range over a $10\,$year runtime is shown by the red vertical lines. The forecasts for the combined standard siren and GW weak lensing constraints using a population of BSS only and a range of values for the instrumental $d_L$ uncertainty from $1\%-10\%$ (green lines) are in broad agreement with those presented in Ref.\;\cite{BNS_DETECTIONS}. The combination of these populations of BSS with a population of DSS with $N_{\rm DSS} = 100\,N_{\rm BSS}$ and an instrumental uncertainty of $10\%$ (blue lines) demonstrates the importance of including the large number of noisier, but higher redshift, sources. The blue lines stop at $\bar{n}_{\rm BSS} = 10\,$deg$^{-2}$ as DSS source numbers greater than $10^{3}\,$deg$^{-2}$ are unlikely, indicated by the LVK inferred limits of all BNS mergers. For expected numbers of sources from 3G GW detectors, by combining BSS and DSS observations, constraints on $w_0$ and $w_a$ will improve upon those by the Euclid $3\times2$pt analysis \cite{Euclid_Forecasts} by a factor of $3-10$. The Vera C. Rubin Observatory predicts similar constraints \cite{LSST,Rubin}, therefore a joint standard siren+GW-WL analysis using 3G GW detectors will improve further dark energy constraints from future galaxy surveys. Expected forecasts from 3G detectors for all parameters can be found in Appendix\;\ref{app:forecasts}. 

Fig.\;\ref{fig:w0wa} shows how increasing the detector uncertainty of the BSS sources from $1\%$ to $10\%$  in the combined BSS+DSS constraints (blue lines) increases these combined constraints by a factor of 2, indicating that the forecasts are sensitive to the assumed instrumental error. However, there is still a significant gain in information when combining the two merger populations, demonstrated by the difference between the green and blue lines. These forecasts are also sensitive to the observed distribution of mergers and hence on the choice of population synthesis code used, in this case \texttt{BPASS} \cite{BPASS}. It has been shown how the choice of the star formation rate and metallicity evolution used by the population synthesis code can have a large impact on the distribution of BBH mergers, less so on the BNS merger distribution \cite{filippo}. Forecasts were also produced using different $p(z)$ assumptions to test the sensitivity of our results to this choice, see Appendix \ref{app:pobs}. As expected different redshift distributions can alter the parameter degeneracies and resulting constraints, but this effect is marginal, and does not affect the main conclusions of the paper.

\subsection{2050's and beyond: DeciHz GW detectors}

\begin{figure}
    \centering
    \includegraphics[width=8.6cm]{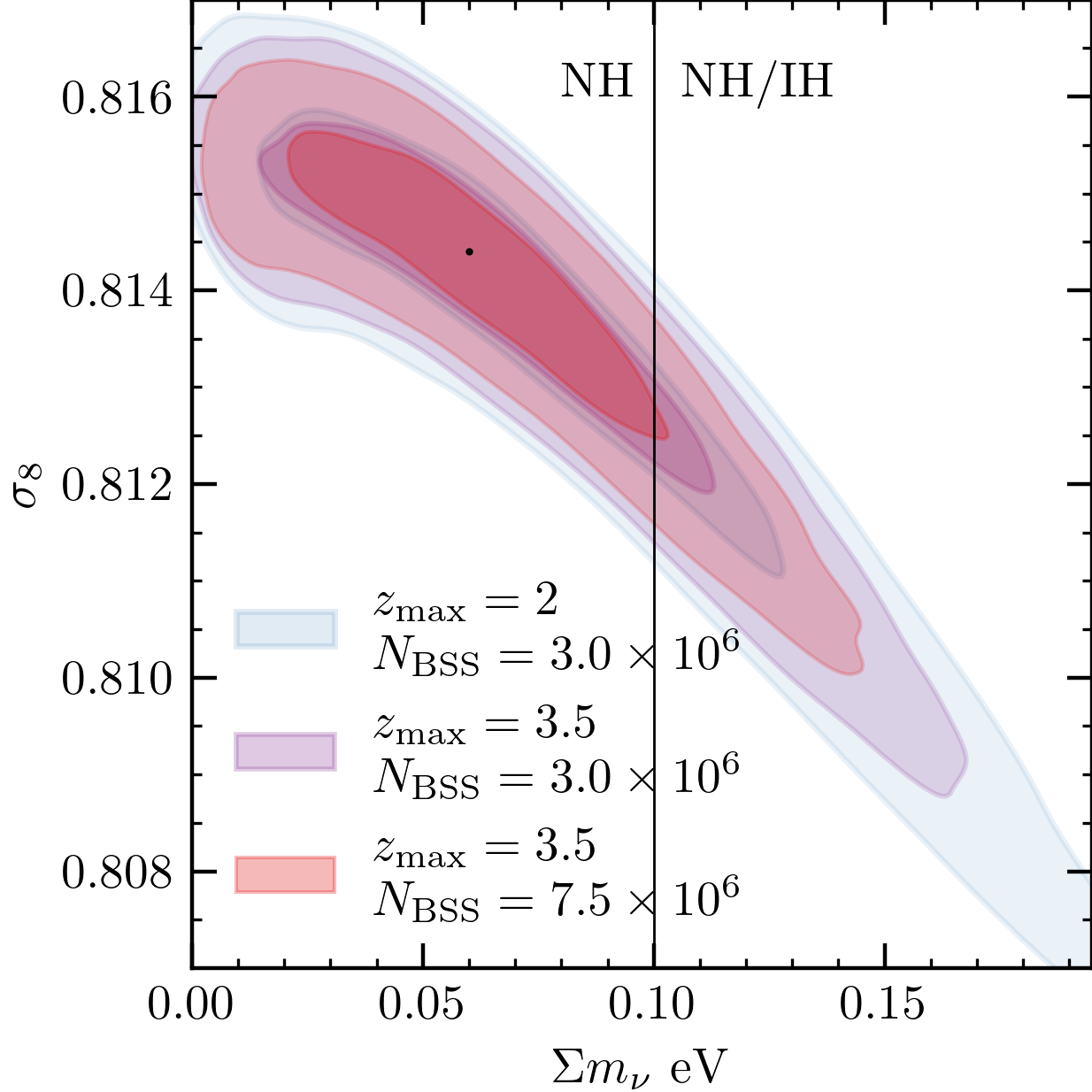}
    \caption{Constraints in the sum of neutrino masses vs $\sigma_8$ plane in the $\nu\Lambda$CDM cosmological model (see Tab.\;\ref{tab:parameters}) from a joint gravitational wave weak lensing+standard siren analysis. $z_{\rm max}=2$ corresponds to merging Binary Neutron Stars with an observed electromagnetic (EM) transient counterpart (spec$-z$), observed up to a maximum redshift of $2$.  $z_{\rm max}=3.5$ is a more optimistic population with spectroscopic redshifts up to $3.5$, and the inclusion of sources without an EM counterpart (merging black holes or black hole---neutron stars) whose host galaxy has been identified through exquisite sky localisation of the gravitational wave (GW) signal. This is only possible with DeciHz GW detectors. Here it is assumed the instrumental uncertainty $\sigma_{d_L} / d_L = 0.4\%$ \cite{ULTRA}. The region left of the black vertical can only correspond to the normal neutrino mass hierarchy (NH), while that to the right can correspond to either the normal or inverted hierarchy (NH/IH).}
    \label{fig:nuLCDM}
\end{figure}

DeciHz detectors will observe the inspiral of orbiting binaries for months to years before merger as the satellites orbit the Sun. The resulting precise $d_L$ determination and sky localisation will allow identification to a host galaxy for many sources, without the need for an EM transient counterpart \cite{ULTRA,Beyond_Lisa,DECIHZ_LINK,DeciHz_new}. This greatly increases the expected number of BSS, comparable to the number of DSS.

\begin{figure*}
    \centering
    \includegraphics[width=17.8cm]{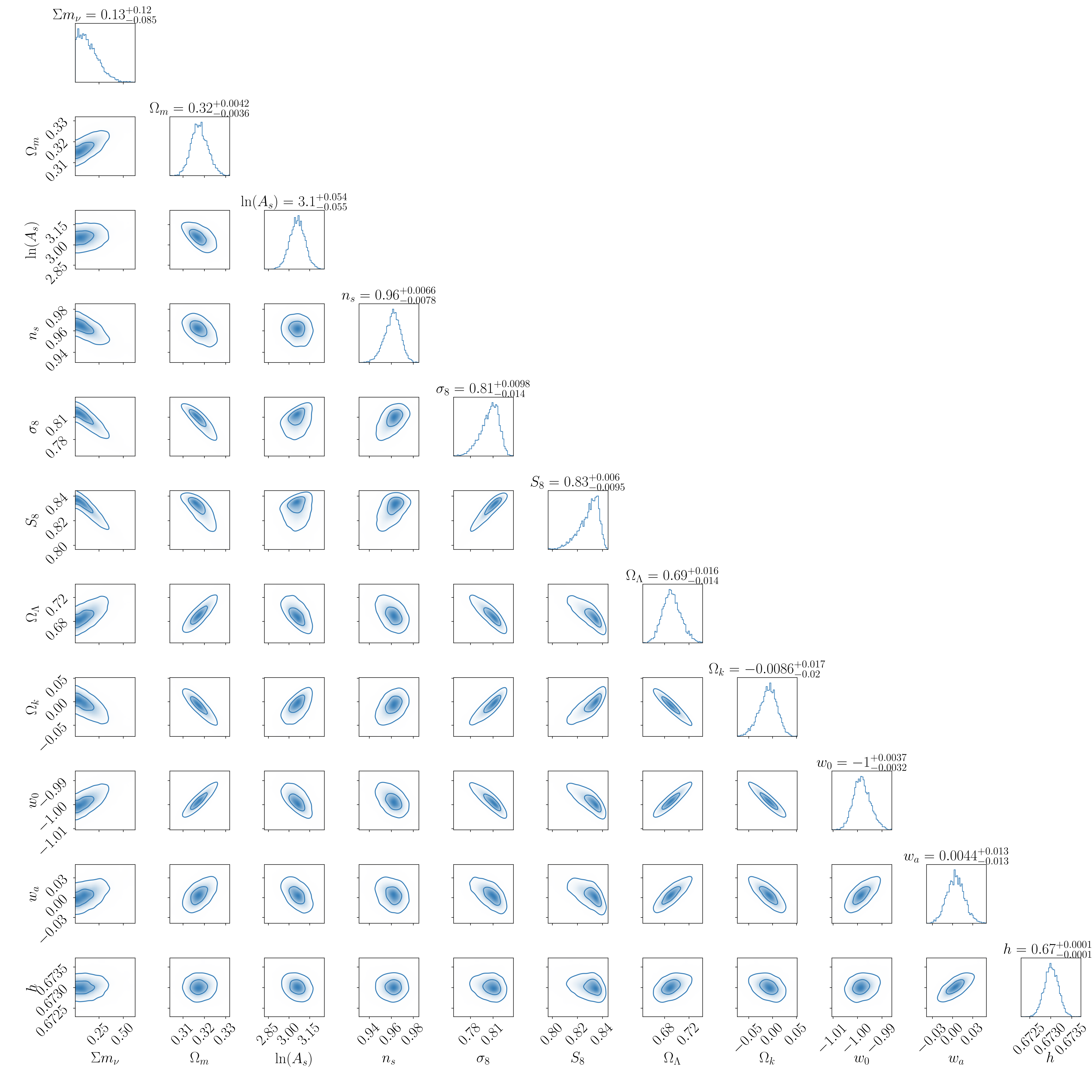}
    \caption{Constraints on all cosmological parameters in the $\nu kw$CDM cosmological model (see Tab.\;\ref{tab:parameters}) from the joint inference analysis of gravitational waves (GWs). This involves combining a standard siren distance measurement with a weak lensing analysis of the GWs. Here we use $3\times10^{6}$ sources up to $z_{\rm max}=2$ with $\sigma_{\rm GW}/d_{L}=1\%$. Created using the \texttt{corner} package \cite{corner}.}
    \label{fig:nuwkCDM}
\end{figure*}

In Fig.\;\ref{fig:w0wa}, when we extend to source numbers only possible with DeciHz detectors the precision will be significantly improved (for example see Tab.\;\ref{tab:DeciHzconstraints}). This huge population of sources with accurate redshifts allows weakly lensed GWs to be used for more challenging statistical cosmological tests. The value of the sum of neutrino masses $\Sigma m_{\nu}$ is presently only constrained by upper bounds. The most stringent constraint is $\Sigma m_{\nu} < 0.12\,$eV ($95\%$ C.L.) coming from Planck 2018 including lensing + BAO \cite{Planck}. The value of $\Sigma m_{\nu}$ impacts structure formation, and  precise determinations could shed light on whether neutrinos follow a normal or inverted hierarchy \cite{mnu_Planck}, adding valuable information to the question of whether neutrinos are Dirac or Majorana fermions \cite{mnu_hier}. The normal hierarchy predicts $\Sigma m_{\nu} > 0.06\,$eV, while in the inverted hierarchy scenario $\Sigma m_{\nu} > 0.10\,$eV, so they can be distinguished in the case where neutrinos follow a normal hierarchy and $\Sigma m_{\nu} < 0.1\,$eV.

It is interesting to investigate how informative the weak lensing of GWs can be on the sum of neutrino masses. Refs\;\cite{ET_mnu,GWmnu,GWmnu2} explored how combining a standard siren measurement of GWs with other probes (CMB, Supernovae and BAO) could be used to improve constraints on the sum of neutrino masses, but as far as we are aware our work demonstrates the first use of GWs from merging binaries as an independent probe of the sum of neutrino masses. We include a more optimistic source distribution where BSS are comprised of highly localised BNS, BBH and BHNS mergers and have an associated spectroscopic redshift up to $z_{\rm max}=3.5$, similar to that used in \cite{ULTRA} where they assumed $z_{\rm max} = 5$. Fig.\;\ref{fig:nuLCDM} shows constraints in $\nu\Lambda$CDM in the $\Sigma m_{\nu}$ vs $\sigma_8$ plane, which can only be investigated with the added lensing information and so represents an entirely novel use of GWs. We see the degeneracy slope between these two parameters, where the free-streaming of higher mass neutrinos impedes clustering. Because of the improved $d_L$ measurement in DeciHz detectors we set $\sigma_{\rm GW} / d_L=0.4\%$ based on the median BBH distance uncertainty at $z\!\sim\!1.5$ of Ref.\;\cite{ULTRA}. More conservative results can be seen in Appendix\;\ref{app:decidl}. In Fig.\;\ref{fig:nuLCDM} the results for $z_{\rm max}=2$ correspond to the same BSS population as in Fig.\;\ref{fig:wCDM_3G}. With the higher redshift population and larger source numbers, a $\sim\!1\sigma$ preference for NH can be determined with $\sigma(\Sigma m_{\nu}) = 0.04\,$eV. $\sigma(\Sigma m_{\nu}) = 0.05\,$eV is still achievable with the same population and a much smaller source number, even when increasing the distance uncertainty to $1\%$. Assuming the smaller number and lower redshift scenario with $\sigma_{\rm GW}/d_{L}=1\%$ gives $\sigma(\Sigma m_{\nu}) = 0.07\,$eV. Due to cosmic variance and the assumed $f_{\rm sky}$, improvement with number density diminishes past $\bar{n} = 500\,$deg$^{-2}$.

The most cosmology agnostic model used in our framework is the $\nu k w$CDM model. Allowing both curvature and dark energy to vary changes some parameter degeneracies \cite{kw}. Fig.\;\ref{fig:nuwkCDM} shows contours for all parameters, in this case there is $\bar{n}_{\rm BSS} = 200\,$deg$^{-2}$ ($N=3\times10^{6}$) with $\sigma_{\rm GW} / d_L = 1\%$. Even in this case we obtain percent or sub-percent accuracy on all parameters, besides those relating to neutrinos. This demonstrates the power of this joint standard siren+GW-WL method. Forecasts for constraints on all parameters in each cosmological model from DeciHz detectors can be found in Appendix\;\ref{app:forecasts}.

\subsection{\label{sec:var} Cosmologically varying \texorpdfstring{$p(z)$}{p(z)}}

\begin{figure}
    \centering
    \includegraphics[width=8.6cm]{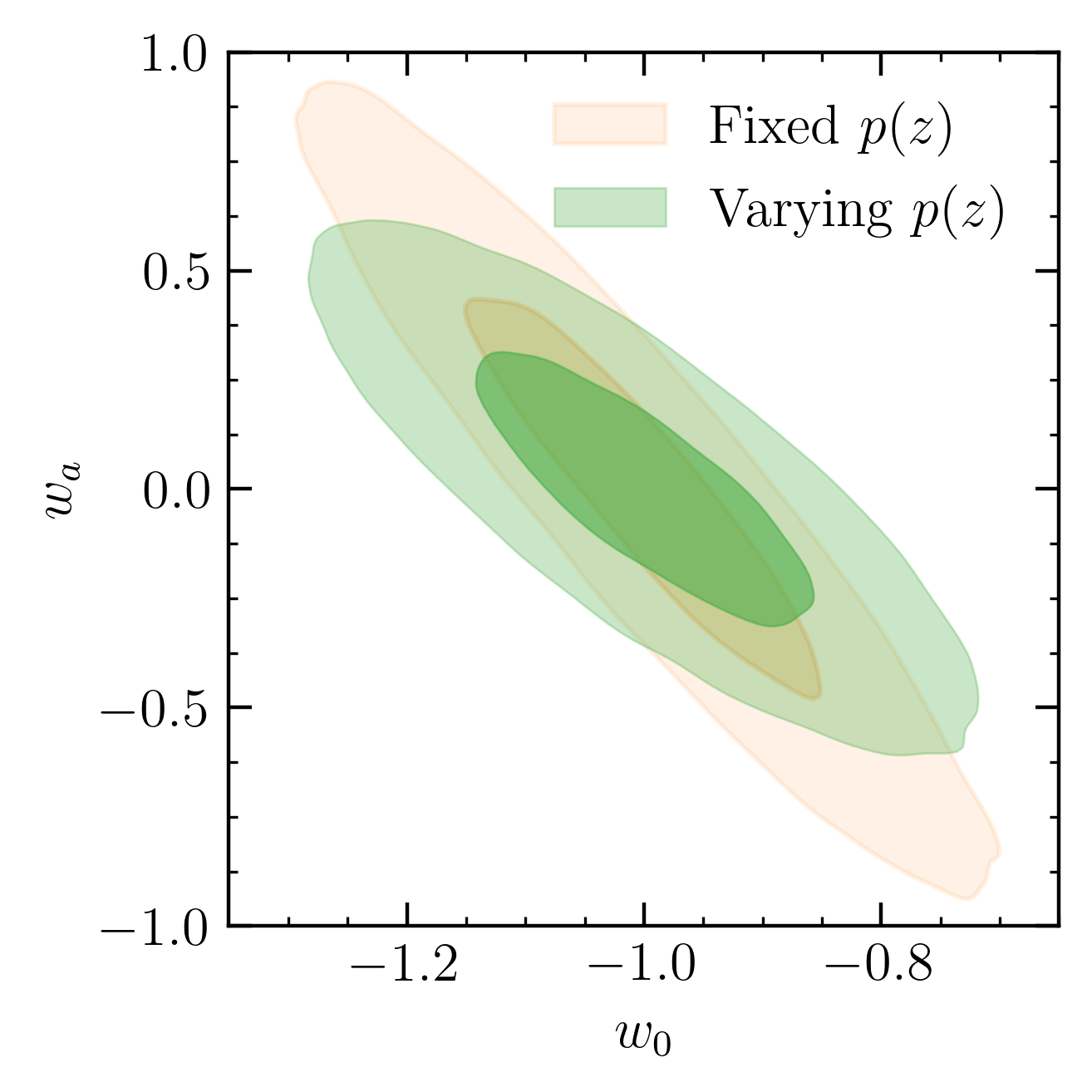}
    \caption{Constraints on dark energy equation of state parameters $w_0$ and $w_a$ in the $w$CDM cosmological model (see Tab.\;\ref{tab:parameters}) from the weak lensing of gravitational waves only. The orange contour is the situation of a fixed, observed $p(z)$ in the weak lensing analysis. The green contour is when we use prior knowledge about the expected cosmic merger rate density of sources, and then exploit the dependence of $p(z)$ on cosmological parameters through the comoving volume.}
    \label{fig:varpz}
\end{figure}

For the weak lensing of gravitational waves, in this framework we generate the observed redshift distribution using a fixed $\mathcal{R}(z)$ in Eq.\;(\ref{eq:pth}). If the form of $\mathcal{R}(z)$ is well-known, then the cosmology dependence of $p(z)$, through the comoving volume term, can be exploited. Expected astrophysical parameters such as spin and mass distributions contributing to the form of $\mathcal{R}(z)$  can be found from many observations, and through this added knowledge we can perform a first order decoupling of astrophysics and cosmology. This will be possible with 3G detectors as we start to constrain the formation channels of binary mergers, leading to an observationally motivated expression for $\mathcal{R}(z)$  \cite{future_Rz_A,future_Rz_B,BBH_Rz_errors}. Knowing $\mathcal{R}(z)$ means the cosmology dependence of $p(z)$ can be exploited, and also reduces our sensitivity to selection effects usually present in an observed $p(z)$. Again we use \texttt{BPASS} for $\mathcal{R}(z)$'s. We choose this non-parametric model due to its good agreement with observations. A parametric model, while possible to marginalise over, could produce wildly incorrect rates and therefore not be a useful model. Fig.\;\ref{fig:varpz} shows a comparison between forecasts found when $p(z)$ is kept fixed with the fiducial cosmology, or allowed to vary in the derivative calculation. These contours show constraints given by the weak lensing of GWs only (not combined with the standard siren measurement). It can be seen how the $w_0-w_a$ degeneracy is shifted, and information is gained on $w_a$. For other parameters, allowing $p(z)$ to vary leads to some cancellation of parameter dependence and therefore weaker constraints. The combined standard siren and weak lensing constraints are only marginally affected. Tests were performed to check whether uncertainties in $\mathcal{R}(z)$ could affect parameter uncertainties, and we found that even large errors in $\mathcal{R}(z)$ do not have an effect on quoted results.

\section{Discussion and Conclusions}
\label{sec:concl}

Future $3^{\rm rd}$ generation gravitational wave detectors will observe huge numbers of binary mergers, sufficient for challenging statistical cosmological tests such as weak lensing. Two confirmed detector proposals, the Einstein Telescope and Cosmic Explorer plan to come online in the late 2030's - early 2040's. Looking further, there will be an incredible boom for GW cosmology with the advent of DeciHz observatories, but due to the technical difficulty their status will not be clear until after the pioneering space-based GW observatory LISA has been launched, also in the 2030's. Therefore, as we consider 10 years of data taking, these forecasts are for far in the future, when GW cosmology has become a tried and tested method. Here we explore those cosmological parameters that are presently of interest, such as the dark energy equation of state parameters and the sum of neutrino masses, to illustrate the power of the method and compare with other surveys which provide predictions for the same parameters. 

We find GW weak lensing tomography is set to be a valuable cosmological tool for 3G detectors. Bright Standard Sirens found from multimessenger gravitational wave and electromagnetic observations could constrain the dark energy equation of state parameters $w_0$ and $w_a$ with comparable precision to Euclid, depending on the number of sources and their uncertainty. It is likely that true numbers of BSS will be similar to the upper limits quoted in Ref.\;\cite{BNS_DETECTIONS} due to the existence of other GW (ET, LISA) and electromagnetic (Fermi-GBM, Swift-BAT etc.) observatories. In the case where we combine a population of BSS with realistic numbers of Dark Standard Sirens  coming from binary black hole mergers, constraints on $w_0$ and $w_a$ improve upon Euclid, and give comparable or better constraints to forecasts for a CMB-S4-like experiment with $10^{9}$ spectrometer hours \cite{mmIm}, and a 1024 dish HIRAX experiment which expects a precision $\sim\!$ few percent \cite{HIRAX}. Therefore it is vital the tools necessary for statistical inference of DSS redshifts are developed. The situation could be improved even further by high redshift spectroscopic galaxy surveys such as MegaMapper \cite{MM}, which could allow a spec$-z$ statistical redshift determination for DSS mergers. The advantage of GW-WL over intensity mapping is we have a much better understanding of the physics underlying how GWs are lensed (essentially just General Relativity), compared to the complex relationship between the distribution of, for example, neutral Hydrogen and dark matter. The advantage over a CMB experiment is the lack of foregrounds confusing the GW signal. A final benefit of GW-WL is the ability to use an astrophysically motivated cosmic merger rate density of sources. Large numbers of future observations will help to constrain the formation channels of binary mergers, allowing us to exploit the cosmology dependence of $p(z)$. We find allowing $p(z)$ to vary with cosmology leads to improved constraints on the time-variation dark energy, but degrades other geometry parameters such as $h$ and $\Omega_{\rm m}$. The main difficulty to contend with for GWs is observing a signal over the background with a high enough SNR to accurately measure its parameters, such as distance and chirp mass. 

For space-based DeciHz GW detectors, many more BSS observations are expected due to the incredible sky localisation of these detectors, including DECIGO and the Big Bang Observatory. There will be exquisite precision on geometry parameters due to very large number of $d_L - z$ data points with small $d_L$ errors. These detectors could achieve a $1\sigma$ preference for the neutrino normal hierarchy using a fractional sky coverage of $f_{\rm sky} = 0.36$.  While not significant enough as a single probe to determine the neutrino mass hierarchy, these results demonstrate for the first time how the weak lensing of GWs could provide valuable extra information on the sum of neutrino masses when combined with future CMB experiments \cite{CMBHD} and HI intensity mapping surveys \cite{SM_LSS}. For these results to be possible, the substantial challenge to overcome is building and deploying the space-based DeciHz GW detectors into their heliocentric orbits. Another recently explored possibility is a Lunar-based DeciHz GW detector, which would be easier to maintain and have comparable sensitivity to the proposed space-based DeciHz detectors \cite{GLOC}.

A limitation of this study is using single values for uncertainties for the whole population of GW sources. A natural extension to this work is to produce a realistic set of GW measurements through a Fisher forecast of their parameter uncertainties, then perform the $d_L-z$ and power spectrum fitting on this realistic set of measurements.

\begin{acknowledgments}

CTM thanks Danny Laghi for helpful discussion on statistical redshift inference, and Suvodip Mukherjee for useful exchange on the shot noise modification due to the gravitational wave localisation uncertainty. CTM is supported by a Science and Technology Facilities Council (STFC) Studentship Grant. AT is supported by an STFC Consolidated Grant. The \texttt{Python} packages \texttt{numpy}, \texttt{scipy}, \texttt{numdifftools}, \texttt{matplotlib} and \texttt{corner} have been used in this work. For the purpose of open access, the authors have applied a Creative Commons Attribution (CC BY) licence to any Author Accepted Manuscript version arising from this submission.
\end{acknowledgments}

\appendix

\section{Choice of GW source distributions \label{app:pz}}

\subsection{Selection Functions \label{app:sel}}
Here assumptions on our survey selection functions are varied to test the sensitivity of results to these choices, which are based on predictions for forthcoming electromagnetic and gravitational wave detectors. The galaxy survey selection function $f(z)$, seen in Eq.\;(\ref{eq:fz}), depends only on the pivot redshift $z_{\rm pivot}$. Instead of changing the form, we investigate the impact of the choice of this pivot redshift in the BSS case. It is found that, because the bulk of the sources are at lower redshifts that are accessible to spectroscopic surveys, varying the pivot redshift and hence the number of sources in the high redshift tail has a small impact on results.

For the GW selection function, we modify the power in the exponential in Eq.\;(\ref{eq:gz}) and the value of $r_{\rm cut}$. Varying the third power to a quadratic or quartic term has a very small effect as the change in the shape of the distribution leads to less sources at lower redshifts but more at higher redshifts. Changing the value of $r_{\rm cut}$ has the largest impact on the presented results, as we are essentially changing the sensitivity of our GW detector. Decreasing $r_{\rm cut}$ by $1\,$Gpc degrades forecasts by $\sim\!10\%$ without affecting the parameter degeneracies. The general conclusions of the paper including the benefits of combining BSS and DSS, and a standard siren+GW-WL analysis producing competitive forecasts with other future probes, are not affected.

\subsection{Observed \texorpdfstring{$p(z)$}{p(z)} \label{app:pobs}}

\begin{figure}
    \centering
    \includegraphics[width=8.6cm]{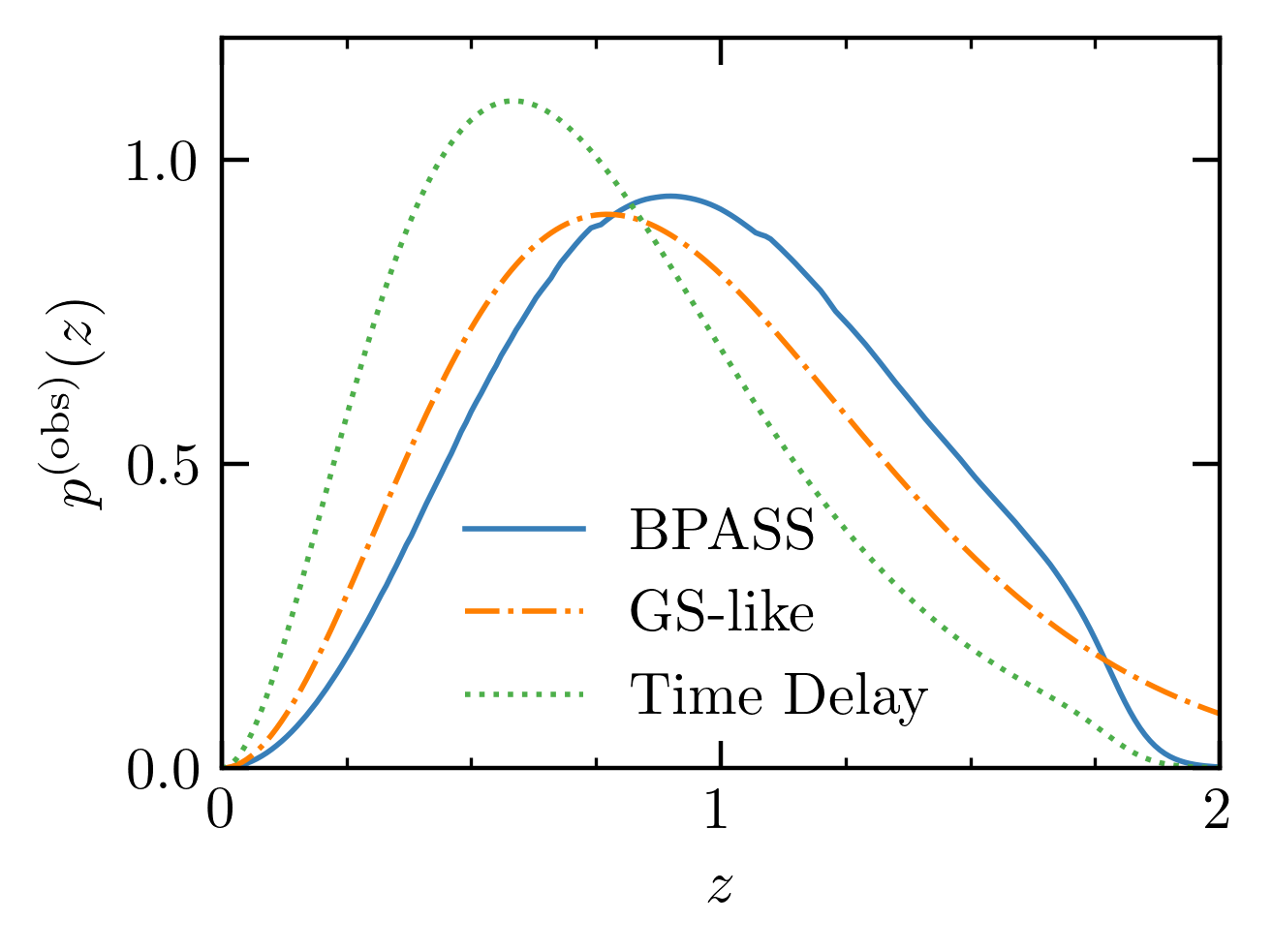}
    \caption{Alternative observed redshift distributions to those used in the main analysis (\texttt{BPASS}, a population synthesis code). `GS-like' is a galaxy-survey like redshift distribution. The last model assumes binary populations follow the star formation rate, with some time delay between the formation and merger.}
    \label{fig:pz_alt}
\end{figure}

\begin{figure}
    \centering
    \includegraphics[width=8.6cm]{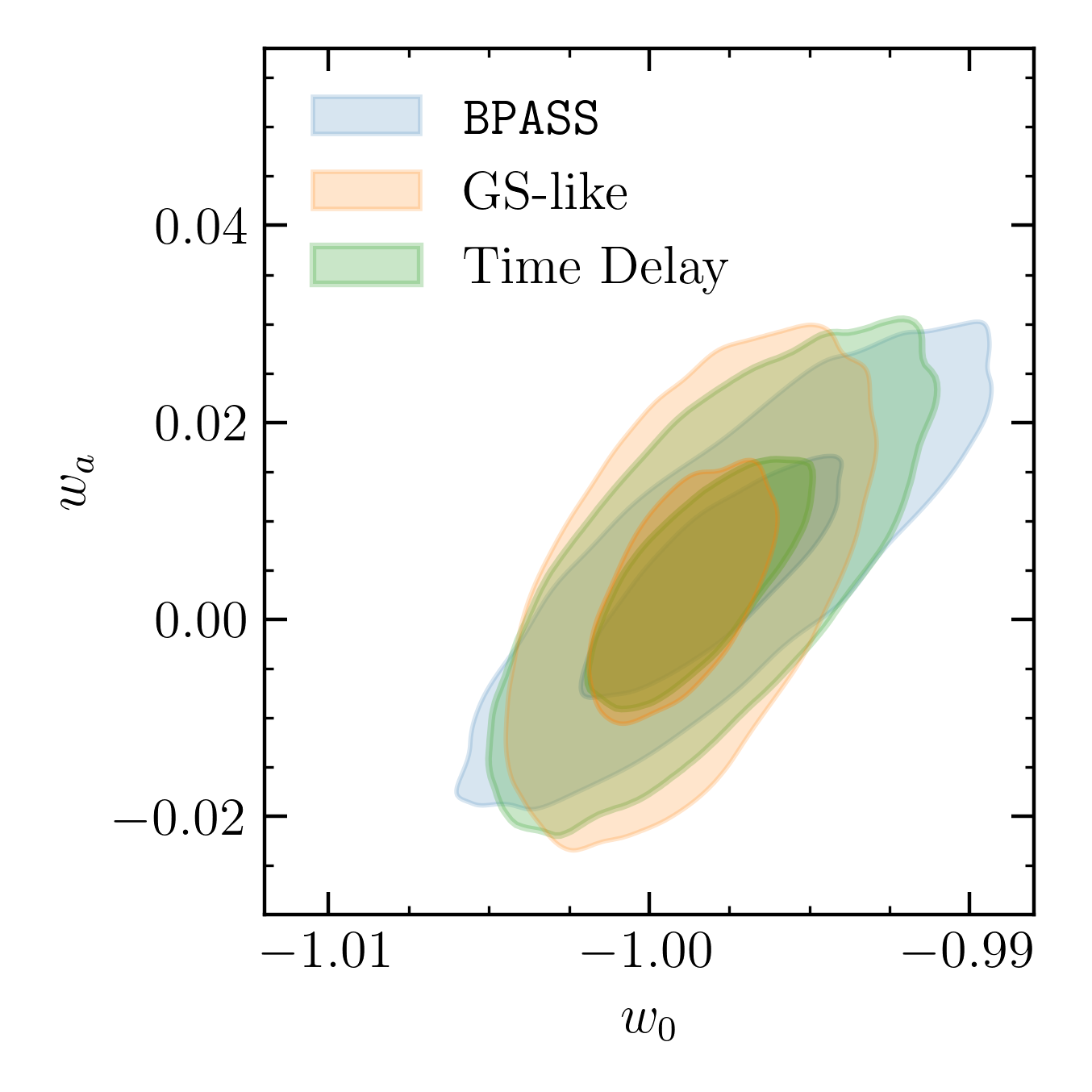}
    \caption{Contours comparing forecasts in the $w_0-w_a$ plane using three different assumptions on the observed redshift distribution of binary mergers. Either a \texttt{BPASS} $\mathcal{R}(z)$, a redshift distribution using a Delay Time Distribution between the SFR and the merger, or a galaxy-survey like (GS-like) distribution.}
    \label{fig:pz_comp}
\end{figure}

Different choices for the observed $p(z)$ can be made. We can use a shifted galaxy survey $p(z)$, 
\begin{equation}
    p(z) \propto z^{2} e^{-(z/z_0)^{2}} \, ,
\end{equation}
with  $z_0 = 2 / \sqrt{2}$ which is higher than the Euclid value of $z_0 = 0.9/\sqrt{2}$ to reflect the greater redshift range of 3G GW detectors. This is the distribution used in  \hyperlink{cite.GCAT}{CT19}, and will be referred to as galaxy survey-like (GS-like).

Another route is changing the assumption on the cosmic merger rate density $\mathcal{R}(z)$. One way is by using a specific delay time distribution (DTD) between the formation of binaries, which follows the star formation rate (SFR), and the eventual inspiral and merger. This is similar to the \texttt{BPASS} results \cite{BPASS}, without any modelling of the complicated physics of binary inspirals.
\begin{equation}
    \mathcal{R}(z) \propto \int_{0}^{t_{\rm max}} {\rm SFR}(\tau)P(\tau) \, d\tau \, .
\end{equation}
$P(\tau)$ is the DTD, and $t_{\rm max}$ is the age of the Universe minus the lookback time to the galaxy. To test the sensitivity of our results we use the more extreme `Slow' model in Ref.\;\cite{BNS_DTD}. These distributions are seen in Fig.\;\ref{fig:pz_alt}.

Although different assumptions on the source distributions do affect the forecasts, the effect is marginal as demonstrated in Fig.\;\ref{fig:pz_comp}

\section{Numerical Derivatives \label{app:derivatives}}

As discussed in Ref.\;\cite{Euclid_Forecasts} section 4.5.2, great care is needed when finding numerical derivatives of the power spectrum---especially in the case of a tomographic analysis. This is because the response of the power spectrum to a changing cosmological parameter can vary greatly over different $\ell$ modes and redshift bin correlations. The choice of the step size for the derivative becomes non-trivial, and tests are needed to ensure results are not overly sensitive to this choice.

The method adopted here uses the \texttt{numdifftools} package. The Derivative function is an adaptive central difference approximation scheme which can calculate the derivative of $\mathcal{C}^{\kappa\kappa}_{\ell}$ for a range of step sizes, returning the optimal step size for each $\ell$ mode independently. This optimal step can vary greatly over $\ell$ modes and tomographic bin combinations. But using different values of the step size for different $\ell$ modes causes discontinuities in the derivative and an unrealistic forecast. Therefore for each parameter, we use the median of the optimal step sizes of each $\ell$ mode and correlation returned by the Derivative function. This may introduce inaccuracies as too large/small a step is being used in some cases. To test whether this median step is indeed optimal for the derivative of the the whole $\mathcal{C}^{\kappa\kappa}_{\ell}$, we define a merit function. We find the absolute difference between the derivative found using a step size and the previous step. This difference between derivatives is found over all ell modes and bin correlations. Then we find the mean of all of these differences (similar results are obtained if the maximum or median difference is used). For an optimal step size, we expect this merit function to be minimised. This implies successive steps are narrowing in on the true value for the derivative. At too small step sizes the difference between successive steps can vary a lot due to numerical error in the calculation. At too large step sizes, we are not picking up the true shape of the function, and the derivative can vary a lot depending on the smaller scale features.

\begin{figure*}
    \centering
    \includegraphics[width=17.8cm]{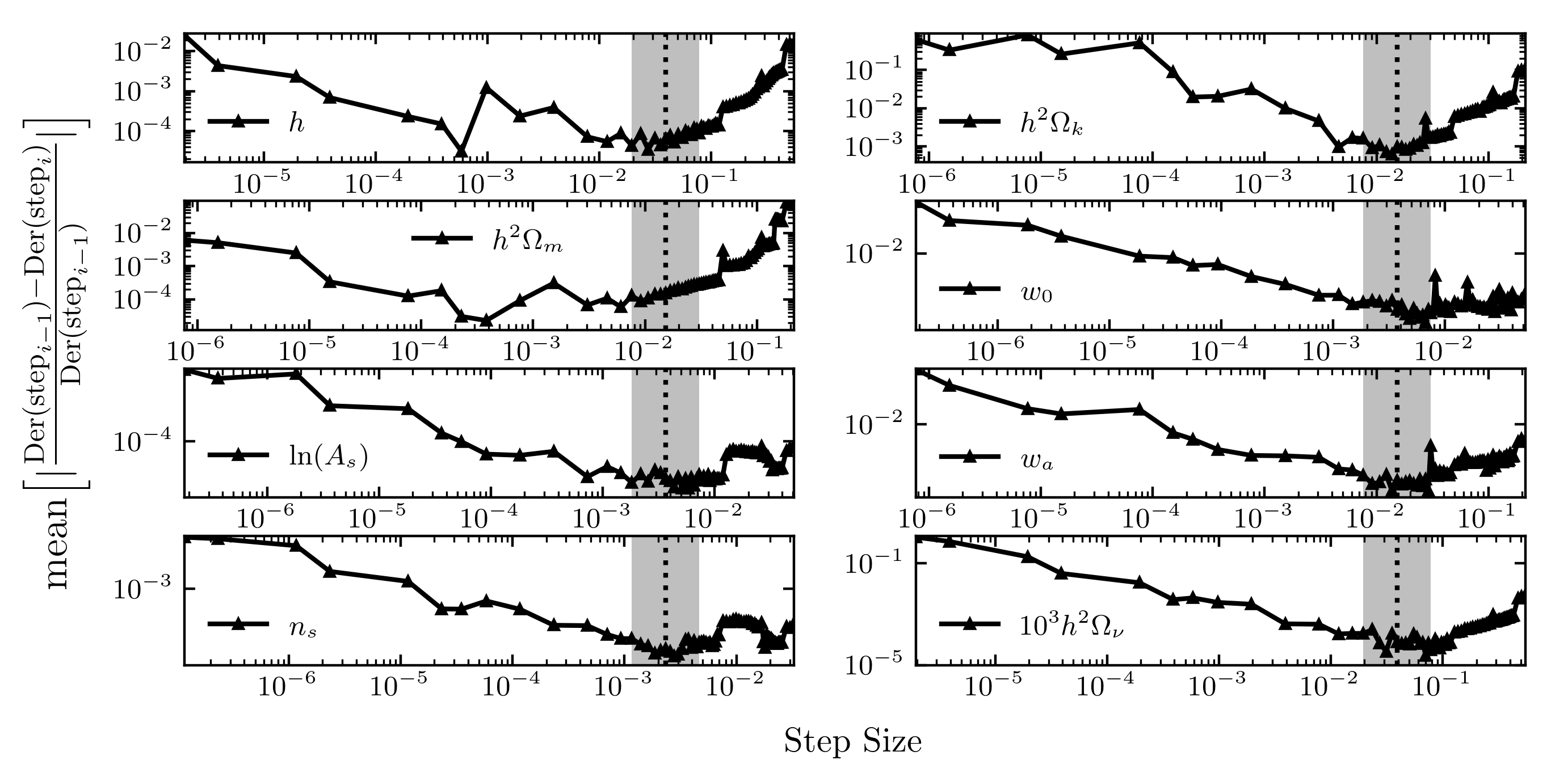}
    \caption{For each parameter a chosen merit function for the derivative of the convergence power spectrum $\mathcal{C}^{\kappa\kappa}_{\ell}$ as a function of the derivative step size is displayed. This is the mean difference between the derivative at a step $i$ and $i-1$. We expect this to be minimised for an optimal step size. Also shown by the vertical black dotted line is the step size found by taking the median of all the `optimal' step sizes returned by the \texttt{numdifftools.Derivative} function for each $\ell$ mode and redshift bin correlation. The grey shaded region shows the step sizes which produce parameter uncertainties within $\sim\!\pm 3\%$ of the uncertainty found using the step size at the dotted line.}
    \label{fig:steps}
\end{figure*}

Once we have checked for optimality using this merit function, stability of results around this step size needs to be tested. Stability was tested for by recalculating forecasts for a large range of step sizes around this median step. We find that the parameter uncertainties vary by at most $\sim\!\pm 3\%$ of the uncertainty found using the optimal step size when the step size is varied in a range of at least $-{\rm log}(2) \leq\,$log(Step / Optimal Step)$\,\leq {\rm log}(2)$ for each parameter. This demonstrates that the results are optimal and robust. These results can be seen in Fig.\;\ref{fig:steps}. Each parameter's merit function has a minimum plateau region, and the optimal step size (vertical dotted line), along with the region where the results are robust (grey shaded region), lie within this minimum plateau. A method of improving the accuracy could be using a different step size for each bin correlation, though from Fig.\;\ref{fig:steps} it is not expected this will have a significant impact on the results presented.

\section{Reducing the value of \texorpdfstring{$\ell_{\rm max}$}{lmax} \label{app:lmax}}

\begin{figure*}
    \centering
    \includegraphics[width=17.8cm]{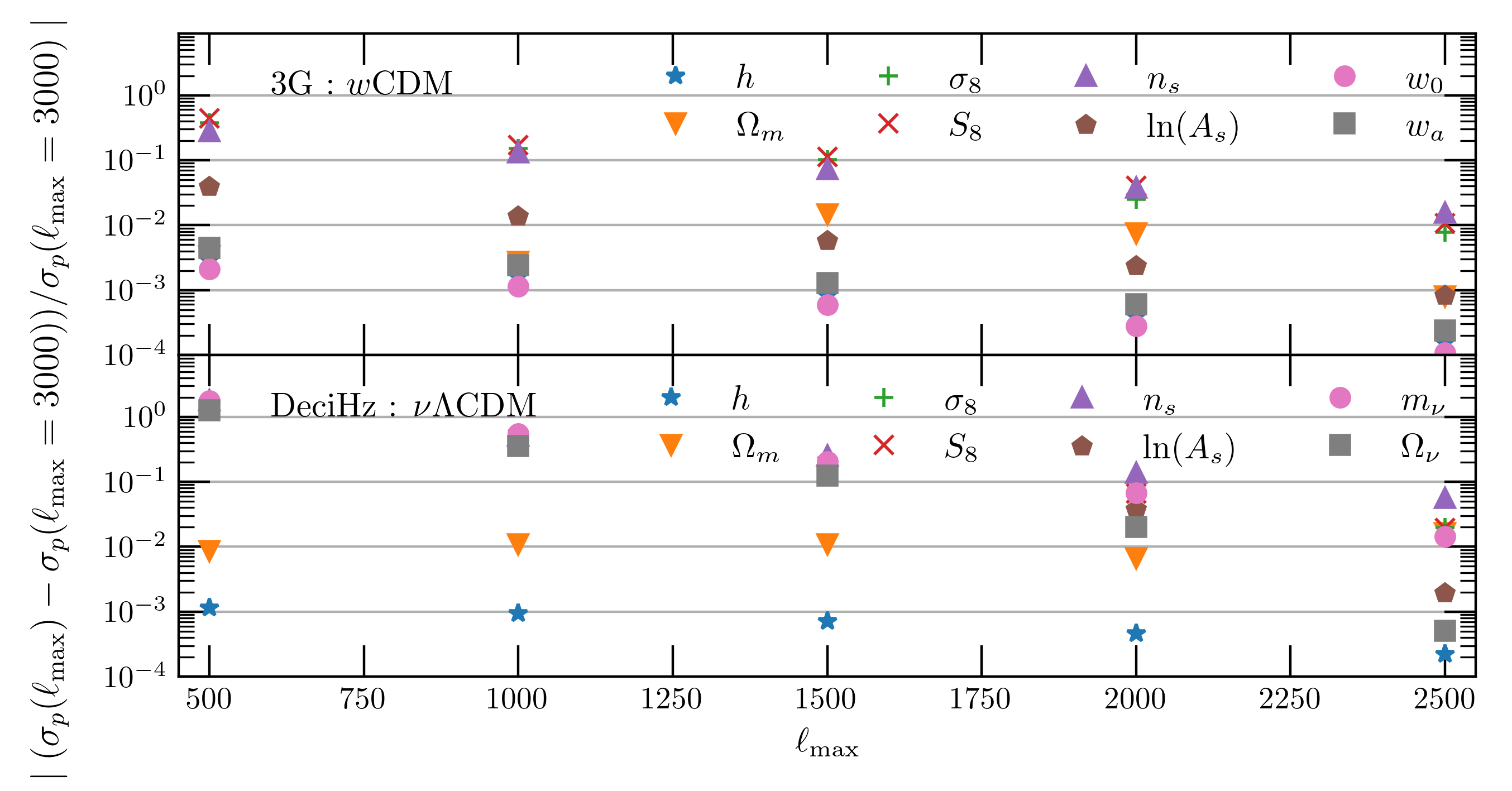}
    \caption{The fractional difference between parameter constraints using a maximum $\ell$ mode of $\ell_{\rm max}=3000$ and other choices of $\ell_{\rm max}$. The comparison is made for the results in Fig.\;\ref{fig:wCDM_3G} (top) and Fig.\;\ref{fig:nuLCDM} (bottom).}
    \label{fig:lmax}
\end{figure*}

The results in Figs.\;\ref{fig:wCDM_3G},\;\ref{fig:nuLCDM} were recreated using different assumptions on the maximum $\ell$ mode. Fig.\;\ref{fig:lmax} shows the fractional difference between parameter constraints found using $\ell_{\rm max}=3000$ (as in the main results) and other choices of $\ell_{\rm max}$. It can be seen that for the dark energy equation of state constrained using 3G GW detectors, the largest difference is only a few tenths of a percent of the original value. The difference is more significant for those parameters constrained only through GW-WL, and here the maximum difference is around $\times 1.4$. The situation is worse in the DeciHz case constraining the $\nu\Lambda$CDM model, this is due to the importance of non-linear scales in constraining the neutrino mass. Here using $\ell_{\rm max}=500$ increases the errors by a factor of $2$ for the power spectrum parameters.

\section{Full parameter constraints \label{app:forecasts}}

Here we show constraints on all free and derived parameters for each cosmological model, for 3G GW detectors in Table\;\ref{tab:3Gconstraints}, and DeciHz detectors in Table\;\ref{tab:DeciHzconstraints}.

\begin{table*}
\centering
\caption{\label{tab:3Gconstraints}%
$1\sigma$ marginalised uncertainties for parameters in the 3G case.Case A uses $30\,000$ BSS and $3\times 10^6$ DSS. Case B uses $3\,000$ BSS and $3\times 10^5$ DSS. In both cases $\sigma_{\rm GW} / d_L = 2\%$ for BSS and $\sigma_{\rm GW} / d_L = 10\%$ for DSS.}
\begin{ruledtabular}
\begin{tabular}{lllllllllll}
\multirow{2}{*}{\textrm{Parameter}}&
\multicolumn{2}{c}{$\Lambda$CDM}&
\multicolumn{2}{c}{$k\Lambda$CDM}&
\multicolumn{2}{c}{$w$CDM}&
\multicolumn{2}{c}{$\nu$CDM}&
\multicolumn{2}{c}{$\nu kw$CDM}\\
 & A & B & A & B & A & B & A & B & A & B \\
\colrule
$h$ $[\times10^{-3}]$ & 0.22  & 0.69    & 0.49  &  1.5 &  1.4 & 4.4 & 0.22  & 0.69  & 2.8 & 7.0 \\
$\Omega_{\rm m}$ $[\times10^{-3}]$ & 0.70 & 2.2   & 2.1  &  6.6 &  5.7  &  18 & 0.70 & 2.2  & 28 &31\\
ln$(10^{10} A_s)$& 0.057 &0.50    & 0.068  & 0.51  &  0.064 &0.51  & 0.66  & 5.3  &  1.2  & 5.5 \\
$n_s$& 0.069  &0.55   & 0.070  &  0.25  &  0.070 & 0.55  & 0.12  & 1.1  & 0.14 & 1.1\\
$\sigma_{8}$ & 0.034 &0.22    & 0.035  & 0.21  &  0.035  &0.21 &  0.036  & 0.20  & 0.069& 0.23\\
$S_8$  & 0.035 &0.22    &  0.061 & 0.12 & 0.035 &  0.22 & 0.037  &  0.20 & 0.23 &0.25\\
$\Omega_{\Lambda}$ & \ding{55}  &  \ding{55} &  0.014 &  0.045 &  \ding{55} & \ding{55}  & \ding{55}  & \ding{55}  & 0.29 &0.31\\
$\Omega_{\rm K}$ &  \ding{55} &  \ding{55} & 0.0072  &  0.023 & \ding{55}  &  \ding{55} & \ding{55}  & \ding{55} & 0.15 &0.16 \\
$w_0$&  \ding{55} & \ding{55}  & \ding{55}  &  \ding{55} &  0.0054 & 0.017  &  \ding{55} & \ding{55}  & 0.087 &0.15\\
$w_a$ & \ding{55}  & \ding{55}  &  \ding{55} & \ding{55}  &  0.036  &0.11  & \ding{55}  & \ding{55}  & 0.32 &0.61\\
$\Sigma m_{\nu}$ [eV] &  \ding{55} &  \ding{55} & \ding{55}  & \ding{55}  &  \ding{55} &  \ding{55} & 1.4  & 11   & 2.8&11\\
\end{tabular}
\end{ruledtabular}
\end{table*}
\begin{table*}
\centering
\caption{\label{tab:DeciHzconstraints}%
$1\sigma$ marginalised uncertainties for parameters in the DeciHz case assuming a number density of Bright Standard Sirens of $200\,$deg$^{-2}$, $3\times10^{6}$ sources using $f_{\rm sky}=0.36$. The maximum redshift these sources are observed to is $z_{\rm max}=2$. Case A uses $\sigma_{\rm GW} / d_L = 0.4\%$, case B uses $\sigma_{\rm GW} / d_L = 1\%$.}
\begin{ruledtabular}
\begin{tabular}{lllllllllll}
\multirow{2}{*}{\textrm{Parameter}}&
\multicolumn{2}{c}{$\Lambda$CDM}&
\multicolumn{2}{c}{$k\Lambda$CDM}&
\multicolumn{2}{c}{$w$CDM}&
\multicolumn{2}{c}{$\nu$CDM}&
\multicolumn{2}{c}{$\nu kw$CDM}\\
 & A & B & A & B & A & B & A & B & A & B \\
\colrule
$h$ $[\times10^{-3}]$ &  0.022 &  0.028   & 0.045  & 0.061  &  0.11  & 0.16  & 0.022  & 0.028  &  0.12 & 0.17 \\
$\Omega_{\rm m}$ $[\times10^{-3}]$ & 0.098  &  0.12 & 0.59  & 0.68  & 1.1 & 1.4 &0.096   &  0.12 & 2.7 &3.9\\
ln$(10^{10} A_s)$ $[\times10^{-3}]$ & 4.9 & 6.3  & 9.3  & 12  &  9.0 &  12 &  41 & 51  & 42 & 56\\
$n_s$ $[\times10^{-3}]$& 4.1  &  5.4 & 4.1  &  5.5 &  4.2  & 5.5 & 5.1 & 6.7  & 7.5 & 8.9\\
$\sigma_{8}$ $[\times10^{-3}]$& 0.78  & 0.98  & 1.3  & 1.6  & 1.4 & 1.7  & 2.8 & 3.5 & 8.4 &12\\
$S_8$ $[\times10^{-3}]$ &  0.80 &  1.0  &  0.86 & 1.1  &  0.91   & 1.1 & 2.9  & 3.6  & 5.7 &7.9 \\
$\Omega_{\Lambda}$ $[\times10^{-3}]$& \ding{55}  & \ding{55}  & 0.85  &  1.0 &  \ding{55} &  \ding{55} & \ding{55}  & \ding{55}  & 10 & 15\\
$\Omega_{\rm K}$ $[\times10^{-3}]$& \ding{55}  &  \ding{55} & 1.4  & 1.7  &  \ding{55} &  \ding{55} &  \ding{55} & \ding{55} & 13 & 19 \\
$w_0$ $[\times10^{-3}]$&  \ding{55} & \ding{55}  & \ding{55}  & \ding{55}  &   0.43  & 0.57 & \ding{55}  & \ding{55}  & 2.8 & 4.1\\
$w_a$ $[\times10^{-3}]$&  \ding{55} & \ding{55}  & \ding{55}  & \ding{55}  & 5.7  &  7.3  &  \ding{55} & \ding{55}  & 10 & 14\\
$\Sigma m_{\nu}$ [meV] &  \ding{55} & \ding{55}  & \ding{55}  &  \ding{55} & \ding{55}  & \ding{55}  & 57  &  67  & 81 & 100\\
\end{tabular}
\end{ruledtabular}
\end{table*}

\section{DeciHz detector assumptions \label{app:decidl}}

\begin{figure}
    \centering
    \includegraphics[width=8.6cm]{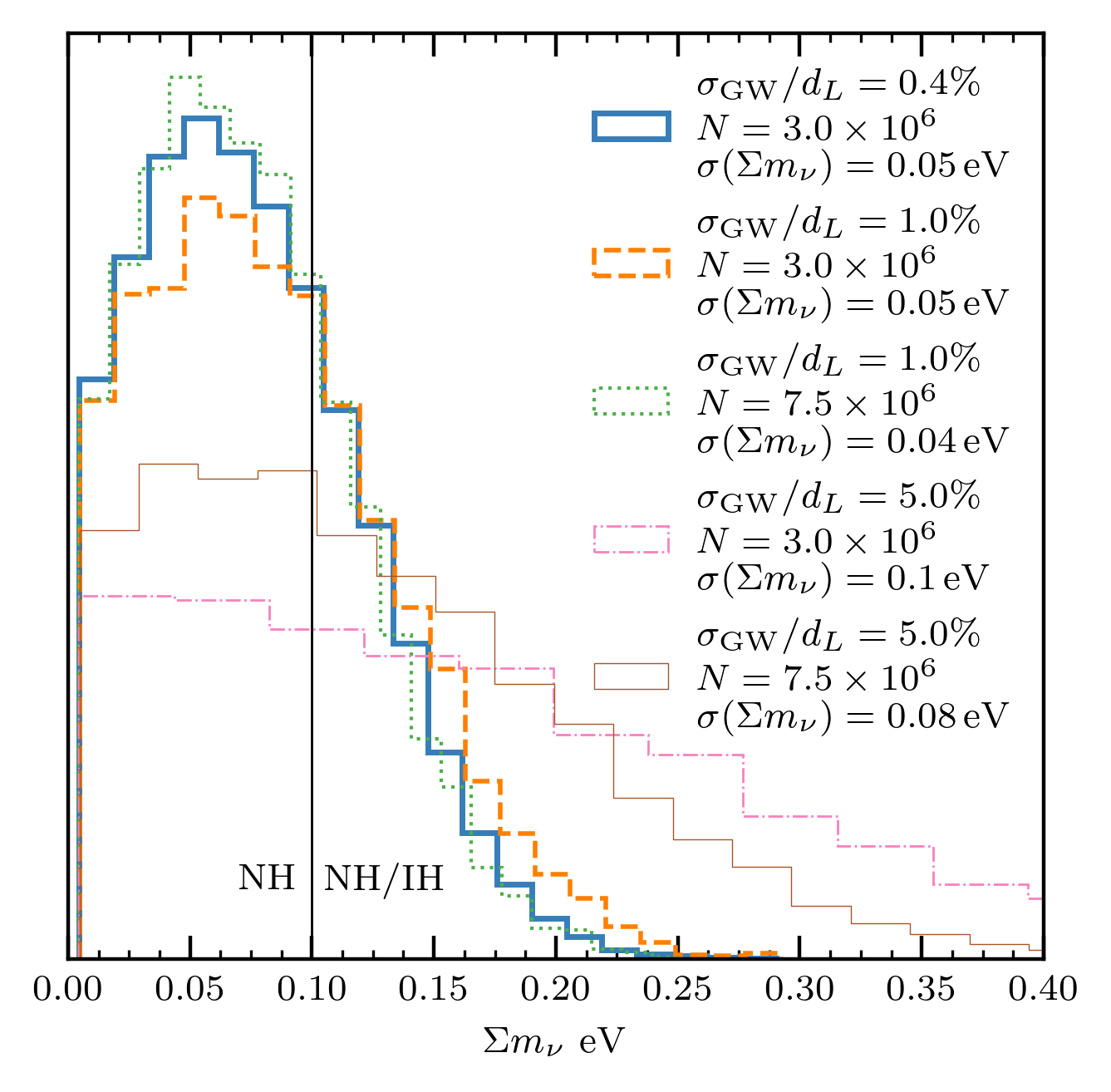}
    \caption{Probability distributions comparing the high redshift population in Fig.\;\ref{fig:nuLCDM} (see Tab.\;\ref{tab:parameters}) with different source numbers and distance uncertainties for DeciHz detectors.}
    \label{fig:dL1}
\end{figure}

In Fig.\;\ref{fig:dL1} we show probability density functions for different assumptions on the instrumental uncertainty and source numbers in DeciHz detectors. In the main analysis we assumed DeciHz detector's very accurate sky localisation would allow well measured luminosity distances (the two measurements are degenerate).

\providecommand{\noopsort}[1]{}\providecommand{\singleletter}[1]{#1}%
\end{document}